\documentclass[longauth]{aa} 

\DeclareUnicodeCharacter{0301}{\'e}
\usepackage{graphicx}
\usepackage{txfonts}
\usepackage{multirow}
\usepackage{hyperref}
\usepackage{color}
\newcommand{\orcid}[1]{\href{https://orcid.org/#1}{\includegraphics[width=10pt]{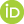}}}

\begin{document} 

    \titlerunning{Diversity of bright long-lived Type~II SNe}
    \authorrunning{T. Nagao et al.}

   \title{Observational diversity of bright long-lived Type II supernovae}

   \author{T. Nagao\inst{1,2,3 \orcid{0000-0002-3933-7861}}\fnmsep\thanks{takashi.nagao@utu.fi}
          \and
          T.~M. Reynolds \inst{1,4,5\orcid{0000-0002-1022-6463}}
          \and
          H. Kuncarayakti \inst{1,6\orcid{0000-0002-1132-1366}}
          \and
          R. Cartier \inst{7\orcid{0000-0003-4553-4033}}
          \and
          S.~Mattila, \inst{1,8\orcid{0000-0001-7497-2994}}
          \and
          K. Maeda \inst{9\orcid{0000-0003-2611-7269}}
          \and
          J. Sollerman \inst{10\orcid{0000-0003-1546-6615}}
          \and
          P. J. Pessi\inst{10\orcid{0000-0002-8041-8559}}
          \and
          J.~P. Anderson \inst{11,12\orcid{0000-0003-0227-3451}}
          \and
          C. Inserra \inst{13\orcid{0000-0002-3968-4409}}
          \and
          T.-W. Chen \inst{14\orcid{0000-0002-1066-6098}}
          \and
          L. Ferrari \inst{15,16\orcid{0009-0000-6303-4169}}
          \and
          M. Fraser \inst{17\orcid{0000-0003-2191-1674}}
          \and
          D. R. Young \inst{18\orcid{0000–0002–1229–2499}}
          \and
          M. Gromadzki \inst{19\orcid{0000-0002-1650-1518}}
          \and
          C. P. Guti\'errez \inst{20,21\orcid{0000-0003-2375-2064}}
          \and
          G. Pignata \inst{22\orcid{0000-0003-0006-0188}}
          \and
          T. E. Müller-Bravo \inst{23,24\orcid{0000-0003-3939-7167}}
           \and
           F. Ragosta \inst{25,26\orcid{0000-0003-2132-3610}}
          \and
          A. Reguitti \inst{27,28\orcid{0000-0003-4254-2724}}
          \and
          S. Moran \inst{29\orcid{0000-0001-5221-0243}}
          \and
          M. González-Bañuelos \inst{21,20\orcid{0009-0006-6238-3598}}
          \and
          M. Kopsacheili \inst{21,20\orcid{0000-0002-3563-819X}}
          \and
          T. Petrushevska \inst{30\orcid{0000-0003-4743-1679}}
          }

   \institute{Department of Physics and Astronomy, University of Turku, FI-20014 Turku, Finland\\
   \email{takashi.nagao@utu.fi}
         \and
         Aalto University Mets\"ahovi Radio Observatory, Mets\"ahovintie 114, 02540 Kylm\"al\"a, Finland
         \and
         Aalto University Department of Electronics and Nanoengineering, P.O. BOX 15500, FI-00076 AALTO, Finland
         \and
         Cosmic Dawn Center (DAWN), Copenhagen, Denmark
         \and
         Niels Bohr Institute, University of Copenhagen, Jagtvej 128, 2200 København N, Denmark
         \and
         Finnish Centre for Astronomy with ESO (FINCA), FI-20014 University of Turku, Finland
         \and
         Centro de Astronom\'ia (CITEVA), Universidad de Antofagasta, Avenida Angamos 601, Antofagasta, Chile
         \and
         School of Sciences, European University Cyprus, Diogenes Street, Engomi, 1516, Nicosia, Cyprus
         \and
         Department of Astronomy, Kyoto University, Kitashirakawa-Oiwake-cho, Sakyo-ku, Kyoto 606-8502, Japan
         \and
         The Oskar Klein Centre, Department of Astronomy, AlbaNova, SE-106 91 Stockholm , Sweden
         \and
         European Southern Observatory, Alonso de C\'ordova 3107, Casilla 19, Santiago, Chile
         \and
         Millennium Institute of Astrophysics MAS, Nuncio Monsenor Sotero Sanz 100, Off. 104, Providencia, Santiago, Chile
         \and
         Cardiff Hub for Astrophysics Research and Technology, School of Physics \& Astronomy, Cardiff University, Queens Buildings, The Parade, Cardiff, CF24 3AA, UK
         \and
         Graduate Institute of Astronomy, National Central University, 300 Jhongda Road, 32001 Jhongli, Taiwan
         \and
         Instituto de Astrofísica de La Plata, CONICET, B1900FWA, La Plata, Argentina
         \and
          Facultad de Ciencias Astronómicas y Geofísicas, Universidad Nacional de La Plata, Paseo del Bosque S/N B1900FWA, La Plata, Argentina
          \and
          School of Physics, University College Dublin, LMI Main Building, Beech Hill Road, Dublin 4, D04 P7W1, Ireland
          \and
          Astrophysics Research Centre, School of Mathematics and Physics, Queen’s University Belfast, Belfast BT7 1NN, UK
          \and
          Astronomical Observatory, University of Warsaw, Al. Ujazdowskie 4, 00-478 Warszawa, Poland
          \and
          Institut d'Estudis Espacials de Catalunya (IEEC), Edifici RDIT, Campus UPC, 08860 Castelldefels (Barcelona), Spain
          \and
          Institute of Space Sciences (ICE, CSIC), Campus UAB, Carrer de Can Magrans, s/n, E-08193 Barcelona, Spain
          \and
          Instituto de Alta Investigación, Universidad de Tarapacá, Casilla 7D, Arica, Chile
          \and
          School of Physics, Trinity College Dublin, The University of Dublin, Dublin 2, Ireland
          \and
          Instituto de Ciencias Exactas y Naturales (ICEN), Universidad Arturo Prat, Chile
          \and
          Dipartimento di Fisica “Ettore Pancini”, Università di Napoli Federico II, Via Cinthia 9, 80126 Naples, Italy
          \and
          INAF - Osservatorio Astronomico di Capodimonte, Via Moiariello 16, I-80131 Naples, Italy
          \and
          INAF – Osservatorio Astronomico di Brera, Via E. Bianchi 46, I-23807 Merate (LC), Italy
          \and
          INAF – Osservatorio Astronomico di Padova, Vicolo dell'Osservatorio 5, I-35122 Padova, Italy
          \and
          School of Physics and Astronomy, University of Leicester, University Road, Leicester LE1 7RH, UK
          \and
          Center for Astrophysics and Cosmology, University of Nova Gorica, Vipavska 11c, 5270 Ajdov\v{s}\v{c}ina, Slovenia
             }

   \date{Received XXX; accepted XXX}

 
  \abstract
   {In various types of supernovae (SNe), strong interaction between the SN ejecta and circumstellar material (CSM) has been reported. This raises questions on their progenitors and mass-loss processes shortly before the explosion. Recently, the bright long-lived Type~II SN 2021irp was proposed to be a standard Type II SN interacting with disk-like CSM. The observational properties suggest that the progenitor was a massive star ($\sim8-18$ M$_{\odot}$) in a binary system and underwent a mass-ejection process due to the binary interaction just before the explosion. Similar scenarios, i.e., a Type II SN interacting with a CSM disk, have also been invoked to explain some Type~IIn SNe.
   }
   {Here, we study the diversity of the observational properties of bright long-lived Type II (21irp-like) SNe. We analyse the diversity of their CSM properties, in order to understand their progenitors and mass-loss mechanisms and their relations with the other types of interacting SNe.
   }
   {We performed photometry, spectroscopy, and/or polarimetry for four 21irp-like SNe. Based on these observations as well as published data of SN~2021irp itself and well-observed bright and long-lived type II SNe including SNe~2010jl, 2015da and 2017hcc, we discuss their CSM characteristics.
   }
   {This sample of SNe shows luminous and long-lived photometric evolution, with some variations in the photometric evolution (from $\sim-17$ to $\sim-20$ absolute mag in the $r$/$o$ band even at $\sim 200$ days after the explosion). They show photospheric spectra characterized mainly by Balmer lines for several hundreds of days, with some variations in the shapes of the lines. They show high polarization with slight variations in the polarization degrees ($\sim1-3$ \% at the brightness peak) with rapid declines with time (from $\sim3-6$ \% before the peak to $\sim1$ \% at $\sim200$ days after the peak).
   The general observational properties are consistent with the disk-CSM-interaction scenario, i.e., typical Type~II SNe interacting with disk-like CSM. At the same time, the variation in the observational properties suggest diversity in the CSM mass and the opening angle of the CSM disk. These variations in the CSM properties are likely to be be related to the binary parameters of the progenitor systems and/or the properties of the progenitor and companion stars.
   }
   {}


   \maketitle
%

\section{Introduction} \label{sec:introduction}

Recent observations have revealed the presence of a large amount of circumstellar material (CSM) around various types of supernovae (SNe), suggesting some unknown extensive mass ejections from their progenitors just before their explosion. For example, the majority of Type~II SNe show signs of substantial CSM interaction at early phases (within a few days after the explosion), including narrow highly ionized/excited lines in the SN spectra \citep[so-called flash-ionization lines;][]{Khazov2016,Yaron2017,Boian2020,Bruch2021} and the rapid rise in their early-phase light curves \citep[LCs; e.g.,][]{Gonzalez2015,Forster2018}. As extreme cases, there are SNe whose main energy source for the radiation is interaction between the SN ejecta and CSM, such as classical Type~IIn SNe \citep[e.g.,][for reviews]{Smith2017, Fraser2020}, some Type~II superluminous SNe \citep[e.g.,][for a review]{Gal-Yam2019}, Type Ia-CSM SNe \citep[e.g.,][]{Uno2023a,Sharma2023}, Type Ibn/Icn SNe \citep[e.g.,][]{Pastorello2008, Hosseinzadeh2017,Pellegrino2022b,Nagao2023a}, and the recently-studied interaction-powered Type II SN~2021irp \citep[][]{Reynolds2025a,Reynolds2025b}. These different types of interacting SNe show diverse photometric and spectroscopic properties. The observational diversity in interacting SNe originates from different SN ejecta properties (e.g., composition, mass, and/or kinetic energy of the SN ejecta) and/or different CSM properties (e.g., composition, total mass, and/or distribution of the CSM). However, we have not yet fully understood the relations between theses SN-ejecta and CSM characteristics and the observational properties of interacting SNe.

Type IIn SNe are bright transients predominantly powered by interaction of the SN ejecta with hydrogen-rich CSM, characterized by strong narrow Balmer lines in their spectra. These narrow lines originate from the ionized/excited unshocked CSM by high-energy photons from the interaction shocks, indicating the low velocity of the CSM \citep[$\lesssim 1000$ km s$^{-1}$; e.g.,][]{Smith2017}. Thus, these narrow lines are regarded as an important sign of the CSM interaction. The more luminous events ($\lesssim -20$ mag) in this category of SNe are often classified as Type~II superluminous SNe \citep[e.g.,][for a review]{Gal-Yam2019}. On the other hand, in some other SNe that are also powered by interaction with hydrogen-rich CSM \citep[e.g., some of the Type~II superluminous SNe, SN~2021irp, and some possibly interaction-powered Type IIL SNe; see][]{Kangas2022,Reynolds2025a} the Balmer lines are dominated by a broad emission component (several thousands of km s$^{-1}$) instead of a narrow one ($\lesssim 1000$ km s$^{-1}$). 

\citet[][]{Reynolds2025b} conducted detailed modeling of the bolometric LC and polarization signals of SN~2021irp, and estimated the CSM mass and distribution as a few M$_{\odot}$ in a disk-like CSM geometry. Based on the estimated configuration of the CSM interaction, they proposed a new interpretation for the origin of the broad-line-dominated Balmer lines to be a torus-shaped photosphere in the SN ejecta, which is locally created by the heating from the hidden ring-shaped interaction region. This demonstrates that the CSM geometry is important to determine the shapes of the Balmer lines in interacting SNe. The authors also discussed the role of the SN ejecta composition in this configuration of the disk-CSM interaction by comparing the shapes of the Balmer lines of SN~2021irp with those of the Type Ia-CSM SN~2020uem \citep[e.g.,][]{Uno2023a,Uno2023b}, which has similar parameters for the CSM interaction as those in SN~2021irp but with Type~Ia SN ejecta ($\sim 1$ M$_{\odot}$ of hydrogen-poor material) instead of Type~II SN ejecta ($\sim 10$ M$_{\odot}$ of hydrogen-rich material). Type~Ia SN ejecta are not dense enough to create an optically thick photosphere at late phases (several hundreds of days after the explosion) in contrast to the case for SN~2021irp. In the configuration of an interaction with disk CSM, the hydrogen-rich gas in the SN ejecta is the key factor to create broad-line-dominated Balmer lines by hiding the narrow lines from the unshocked CSM. In other words, broad-Balmer-line-dominated interacting SNe probably have a Type II SN as the embedded SN.

It is important to study the CSM properties in 21irp-like SNe, i.e., broad-line-dominated and interaction-powered Type II SNe, and the relations between the CSM parameters and the observational properties. This improves our understanding of the radiation processes in interacting SNe, as well as the viewing angle effects for their observational properties. It is also important to understand the progenitor systems and mass-loss mechanisms in 21irp-like SNe, as well as their relations with the other types of interacting SNe. In this paper, we present photometric, spectroscopic and polarimetric observations of four bright long-lived Type~II (21irp-like) SNe (SNe~2018khh, 2021adxl, 2022oo, and 2022mma) including SN~2021irp itself and the well-observed similar events, SNe~2010jl, 2015da and 2017hcc, infer their CSM parameters, and discuss the relations between the CSM parameters and the observational properties. We then discuss their progenitor systems and the mass-loss mechanisms that create their CSM, and their relations with the other types of interacting SNe.

In the following section, we present our sample of SNe with their basic information as well as similar SNe from the literature. In Section~\ref{sec:host}, we discuss the properties of their host galaxies. In Section~\ref{sec:observation}, we present the observational data and the data reduction processes. In Section~\ref{sec:result}, we describe the photometric, spectroscopic and polarimetric properties of the SNe. In Section~\ref{sec:discussion}, we discuss the main energy sources for these SNe, their CSM properties, their possible progenitor systems and mass-loss mechanism, and the relations with the other types of interacting SNe. We conclude the paper in Section~\ref{sec:conclusion}.

\section{SN sample} \label{sec:sn_sample}

We collected observations of several bright long-lived Type~II SNe with broad/intermediate-width Balmer lines, which are similar to SN~2021irp. We also include SN~2021irp itself and well-observed similar events SNe~2010jl, 2015da and 2017hcc in our analysis. The SNe in our sample were arbitrarily picked based on the observational similarity (see Section~\ref{sec:result}) and it is not a complete sample for this kind of SNe. Throughout the paper, we assume the cosmological parameters as $H_0$=73 km s$^{-1}$ Mpc$^{-1}$, $\Omega_M$ = 0.27 and $\Omega_{\Lambda}$ = 0.73 \citep[][]{Spergel2007}, and $R_V$ = 3.1 and the extinction curve by \citet[][]{Cardelli1989}, for extinction. All phases reported in this paper are with respect to the assumed explosion date. The details on the SNe in our sample are described in the following subsections and summarized in Table~\ref{tab:sn_sample}.

\subsection{SN~2018khh}

SN~2018khh was discovered by the All-Sky Automated Survey for Supernovae (ASAS-SN)\footnote{\url{https://www.astronomy.ohio-state.edu/asassn/}} on 20.04 December 2018 UT \citep[58472.04 MJD;][]{Brimacombe2018}. It was classified as a Type~IIn SN about a day after the discovery \citep[][]{Cartier2018}. The redshift of the host galaxy is $z=0.022900$ \citep[][]{Colless2003}, reported in NED, and the corresponding distance modulus is $\mu=34.92$. The last non-detection of the object was on 17.05 December 2018 UT (58469.05 MJD) with the $g$-band magnitude of 17.5 mag, which is about three days before the discovery with the $g$-band magnitude of 16.3 mag \citep[][]{Fremling2022}. We adopt 58470.55 MJD, which is the mid date between the discovery and the last non-detection dates, as the explosion date. 
We estimate the extinction within the host galaxy using the empirical relation between the strengths of Na~I~D interstellar absorption lines and the extinction, derived by \citet[][]{Poznanski2012}. Using the spectrum at Phase~+3.5 days, we measure the equivalent width of Na~I~D lines at the host-galaxy redshift to be $EW\sim0.45$ {\AA}, implying $E(B-V)\sim0.05$ mag for the extinction in the host galaxy. Since the extinction for MW extinction is $E(B-V)=0.021$ mag \citep[][]{Schlafly2011}, we adopt $E(B-V)=0.07$ the total extinction.

\subsection{SN~2021adxl}

SN~2021adxl was discovered by the Zwicky Transient Facility \citep[ZTF;][]{Bellm2019} on 3.54 November 2021 UT \citep[59521.54 MJD;][]{Fremling2021}. It was classified as a Type~IIn SN $\sim 4$ months after the discovery \citep[][]{De2022}. The redshift of the host galaxy was estimated using the narrow emission lines from the host galaxy \citep[$z=0.018$,][]{De2022}, and the corresponding distance modulus is $\mu=34.47$. Since the reported last non-detection was more than $\sim 3$ months before the discovery, we cannot precisely estimate the explosion date. We simply adopt 59500 MJD, which was estimated based on the similarity of the photometric properties between SNe~2021adxl and 2010jl \citep[][]{Salmaso2025}, as the explosion date of SN~2021adxl. Due to the lack of significant Na~I~D interstellar absorption lines at the host-galaxy redshift in our spectra, we ignore the extinction in the host galaxy and only correct for MW extinction \citep[$E(B-V)=0.026$ mag;][]{Schlafly2011}.

\subsection{SN~2022oo}

SN~2022oo was discovered by the Asteroid Terrestrial impact Last Alert System \citep[ATLAS;][]{Tonry2018, Smith2020}\footnote{\url{https://star.pst.qub.ac.uk/sne/atlas4/}} on 12.59 January 2022 UT \citep[59591.59 MJD;][]{Tonry2022}. The first detection by ATLAS of this SN was earlier than the discovery report (59563.66 MJD), and the last non-detection was on 59545.64 MJD, even though this last non-detection (ATLAS-o $> 18.6$ mag in the ATLAS-orange band) was slightly shallower than the first detection (19.1 mag in the ATLAS-orange band). Although we cannot estimate the precise explosion date for this SN, we adopt 59554.65 MJD, which is the mid date between the discovery and the last non-detection dates, as the explosion date. This SN was classified as a Type~II SN about two weeks after the discovery \citep[][]{Pineda2022}. We measured the redshift of the SN or the host galaxy from the positions of the narrow H$\alpha$ lines in the spectra on 20.00/21.00 August 2022 UT ($z=0.091$). The corresponding distance modulus is $\mu=38.01$. Due to the lack of significant Na~I~D interstellar absorption lines at the host-galaxy redshift in our spectra, we ignore the extinction in the host galaxy and only correct for MW extinction \citep[$E(B-V)=0.064$][]{Schlafly2011}.

\subsection{SN~2022mma}

SN~2022mma was discovered by ZTF on 10.21 June 2022 UT \citep[59740.21 MJD;][]{Perez-Fournon2022}. It was classified as a Type~IIn SN within a week after the discovery \citep[][]{Pellegrino2022a}. The redshift of the host galaxy is $z=0.037474$, reported in the NASA/IPAC Extragalactic Database (NED)\footnote{\url{https://ned.ipac.caltech.edu/}}, and the corresponding distance modulus is $\mu=36.00$. The last non-detection of the object was on 8.25 June 2022 UT (59738.25 MJD)  with the $r$-band magnitude of 19.84 mag, which is about two days before the discovery with the $r$-band magnitude of 19.43 mag \citep[][]{Perez-Fournon2022}. We adopt 59739.23 MJD, which is the mid date between the discovery and the last non-detection dates, as the explosion date. Due to the lack of significant Na~I~D interstellar absorption lines at the host-galaxy redshift in our spectra, we ignore the extinction in the host galaxy and only correct for MW extinction \citep[$E(B-V)=0.023$ mag;][]{Schlafly2011}.

\begin{table}
      \caption[]{Summary of the main parameters for the SNe in our sample.}
      \label{tab:sn_sample}
      $
         \begin{array}{lcccc}
            \hline
            \noalign{\smallskip}
            & \rm{Exp.\;date} & \rm{redshift} & \mu & E(B-V)\\
            & (\rm{MJD}) & & (\rm{mag}) & (\rm{mag})\\
            \noalign{\smallskip}
            \hline\hline
            \noalign{\smallskip}
            \rm{SN~2018khh} & 58470.55 & 0.022900 & 34.92 & 0.07\\
            \noalign{\smallskip}
            \rm{SN~2021adxl} & 59500.00 & 0.018 & 34.47 & 0.026\\
            \noalign{\smallskip}
            \rm{SN~2022oo} & 59554.65 & 0.091 & 38.01 & 0.064\\
            \noalign{\smallskip}
            \rm{SN~2022mma} & 59739.23 & 0.037474 & 36.00 & 0.023\\
            \noalign{\smallskip}
            \hline\hline
            \noalign{\smallskip}
            \rm{SN~2010jl} & 55478.50 & 0.010620 & 33.22 & 0.058\\
            \noalign{\smallskip}
            \rm{SN~2015da} & 57030.40 & 0.007085 & 33.63 & 0.98\\
            \noalign{\smallskip}
            \rm{SN~2017hcc} & 58027.40 & 0.0168 & 34.22 & 0.029\\
            \noalign{\smallskip}
            \rm{SN~2021irp} & 59310.30 & 0.0195 & 34.55 & 0.424\\
            \noalign{\smallskip}
            \hline
         \end{array}
         $
         
         \begin{minipage}{.88\hsize}
        \smallskip
        Notes. The explosion date, redshift, and extinction for SNe~2010jl, 2015da, 2017hcc and 2021irp were taken from \citet[][]{Fransson2014,Tartaglia2020,Moran2023,Reynolds2025a}. The distance modulus for SN~2015da was estimated based on the luminosity distance of the host galaxy \citep[$53.2$ Mpc][]{Karachentsev2017}, while the distance moduli for the other SNe are derived from the adopted redshift.
        \end{minipage}
   \end{table}

\section{Host galaxies and local environments} \label{sec:host}

We measured magnitudes for the host galaxies of SNe~2018khh, 2022oo and 2022mma, using the {\sc hostphot}\footnote{\url{https://github.com/temuller/hostphot}} package \citep{MullerBravo2022}. Using utilities available as part of the package, we collected the available optical imaging from the PS1 survey for SNe~2022oo and 2022mma, and DES imaging for SN 2018khh, and performed aperture photometry in the images with a Kron radius optimised by increasing the aperture size until the change in flux is $<0.05\%$. The $g$-band photometry is presented in Table \ref{tab:sn_environment}.

We measured the metallicity of the host galaxies of SNe~2018khh and 2022mma adopting the commonly used N2 technique \citep[][]{Pettini2004}. For this process, we used the spectrum taken by The 2dF Galaxy Redshift Survey \citep[][]{Colless2003} for the host galaxy of SN~2018khh and the one taken by Sloan Digital Sky Survey (SDSS) through SDSS SkyServer \footnote{\url{https://www.skyserver.sdss.org/dr18/}} for the host galaxy of SN~2022mma. The derived value of 12+log(O/H) is 8.75 and 8.63 dex for SNe~2018khh and 2022mma, respectively. From these values, the metallicity is derived to be $Z/Z_{\odot}=1.15$ and $Z/Z_{\odot}=0.87$ for SNe~2018khh and 2022mma, respectively, adopting the solar value to be 12 + log(O/H)$_{\odot}$ = 8.69 dex \citep[][]{Asplund2021}. We note that these values are not for the local environment of the progenitors but for the whole galaxies, and thus the local values can be different. Given that these measurements are biased by the bright central parts of the galaxies, it is likely that the site metallicities are lower.

The main properties of the host galaxies of the SNe in our sample are summarized in Table~\ref{tab:sn_environment}. Many of the hosts are a dwarf spiral galaxies, except those for SNe~2015da, 2018khh and 2022mma. Indeed, the luminosity of the host galaxies are relatively small, compared to the distribution of the SDSS galaxies \citep[see Figure~4 in][]{Tremonti2004}. The hosts of SNe~2010jl \citep[][]{Stoll2011}, 2021irp \citep[][]{Reynolds2025a}, 2021adxl \citep[][]{Brennan2024}, and possibly 2018khh have irregular shapes, which might suggest that they are merging galaxies. In addition, SNe~2010jl and 2021adxl are clearly located on a bright point in the host, which is possibly a star-forming region in the host. The hosts of SNe~2017hcc and 2022oo are so faint that we cannot discuss their shapes. The host of SN~2022mma is relatively bright and thus possibly as massive as usual host galaxies for core-collapse SNe. It is located on a bright point in the outskirts of the host, which is possibly a star-forming region in the host. The host of SN~2015da is a normal spiral galaxy. 

The metallicity of the hosts of the SNe in our sample are also relatively small, compared to the distribution of the SDSS galaxies \citep[see Figure~4 in][]{Tremonti2004}, except the host of SN~2018khh. Since SN~2018khh is located in the outskirts of the irregular-shaped galaxy, the metallicity in the local environment can be smaller than the value estimated for the whole galaxy, which the galaxy core dominates. This trend of hosts with low luminosity and low metallicity is also reported for other luminous Type IIn SNe \citep[][]{Stoll2011}. The star-forming and low metallicity environment might be the key to the origins of this type of SNe.

\begin{table*}
      \caption[]{Summary of the main properties of the environment.}
      \label{tab:sn_environment}
      $
         \begin{array}{lllll}
            \hline
            \noalign{\smallskip}
            & \rm{Host\;galaxy} & \rm{Host\;brightness\;(mag)} & \rm{Metallicity} (Z/Z_{\odot}) & \rm{Data\;sources} \\
            \noalign{\smallskip}
            \hline\hline
            \noalign{\smallskip}
            \rm{SN~2018khh} & \rm{WISEA\;J220315.05-555851.0} & M_{g} = -20.3 & 1.15\;(\rm{global}) &  \rm{This~work, PS1} \\ \hline
            \noalign{\smallskip}
            \rm{SN~2021adxl} & \rm{WISEA\;J114806.88-123841.3} & M_{g}\sim-17.7 & 0.08 \;(\rm{local}) & \text{\citet[][]{Brennan2024}}\\ \hline
            \noalign{\smallskip}
            \rm{SN~2022oo} & - & M_{g}=-18.8 & - &  \rm{This~work, PS1} \\ \hline
            \noalign{\smallskip}
            \rm{SN~2022mma} & \rm{WISEA\;J143901.94+155923.5} & M_{g} = -20.6 & 0.87 \;(\rm{global}) & \rm{This~work, PS1}\\
             & \rm{/SDSS\;J143901.93+155923.1} & & & \\
            \noalign{\smallskip}
            \hline\hline
            \noalign{\smallskip}
            \rm{SN~2010jl} & \rm{UGC\;05189A} & M_{B} = -19.3 & 0.32 \;(\rm{local}) & \text{\citet[][]{Stoll2011}}\\ \hline
            \noalign{\smallskip}
            \rm{SN~2015da} & \rm{NGC\;5337} & M_{B} = -20.69 \pm 0.61 & 0.62 \;(\rm{local}) & \text{\citet[][]{Tartaglia2020}}\\ \hline
            \noalign{\smallskip}
            \rm{SN~2017hcc} & \rm{WISEA\;J000350.27-112828.7} & M_{r} = -16.6 & 0.63 \;(\rm{global}) & \text{\citet[][]{Moran2023}}\\ \hline
            \noalign{\smallskip}
            \rm{SN~2021irp} & \rm{WISEA\;J052327.68+170431.2} & M_{g} = -18.94 & 0.34 \pm 0.05 \;(\rm{global}) & \text{\citet[][]{Reynolds2025a}}\\
            & \rm{/SDSS\;J052327.62+170432.4} & & & \\
            \noalign{\smallskip}
            \hline
         \end{array}
         $
         
         \begin{minipage}{.88\hsize}
        \smallskip
        Notes. The metallicities for SNe~2010jl, 2015da, 2017hcc, 2021adxl, and 2022mma were derived from the value of 12 + log(O/H), adopting the solar value to be 12 + log(O/H)$_{\odot}$ = 8.69 dex \citep[][]{Asplund2021}. The host brightness was corrected for extinction. The metallicities for SNe~2010jl, 2015da and 2021adxl were estimated for their local environments (local), while that for the others is for their host galaxies (global).
        \end{minipage}
   \end{table*}

\section{Observations and data reduction} \label{sec:observation}

We collected the the $r$-band LCs obtained by ZTF through the Automatic Learning for the Rapid Classification of Events (ALeRCE) alert broker \citep[][]{Forster2021}\footnote{\url{https://alerce.science/}} for SNe~2021adxl and 2022mma, and the $o$-band LC obtained by ATLAS for SN~2022oo through the ATLAS Forced Photometry server \citep[][]{Shingles2021}. In addition, we have obtained $g$-, $r$-, $i$-, and $z$-band photometry of SN~2018khh, and the details are shown in Appendix~\ref{sec:app_phot_18khh}. At the same time, we conducted polarimetric and/or spectroscopic observations for SNe 2021adxl, 2022mma, 2022oo, and 2018khh. The details on the polarimetry and spectroscopy are in the following subsections.

\subsection{Polarimetry}

We obtained spectro- and/or imaging polarimetry of SNe 2021adxl and 2022mma using the FOcal Reducer/low dispersion Spectrograph 2 \citep[FORS2;][]{Appenzeller1998} mounted at the Cassegrain focus of the Very Large Telescope (VLT)\footnote{\url{https://www.eso.org/public/teles-instr/paranal-observatory/vlt/}} UT1 telescope at the Paranal observatory and/or the Alhambra Faint Object Spectrograph and Camera (ALFOSC)\footnote{\url{https://www.not.iac.es/instruments/alfosc/}} mounted on the 2.56-m Nordic Optical Telescope (NOT)\footnote{\url{https://www.not.iac.es/}} at the Roque de los Muchachos Observatory. The observing logs are shown in Appendix~\ref{sec:app_obs}.

The spectropolarimetric observations of SN~2021adxl were obtained using the low resolution G300V grism and a half-wave retarder plate (HWP) adopting the optimal set of HWP angles of $0^{\circ}$, $22.5^{\circ}$, $45^{\circ}$ and $67.5^{\circ}$. We analyzed these data with standard methods \citep[e.g.,][]{Patat2006, Nagao2024a}, using IRAF \citep[][]{Tody1986,Tody1993}. After applying cosmic-ray removal, bias subtraction, and flat-field corrections to the object frames, we extracted the ordinary and extraordinary beams of the SN with a fixed aperture size of $10$ pixels. The extracted spectra were rebinned to 50 {\AA} bins to have a better signal-to-noise ratio. We also corrected the polarization angles for HWP zeropoint angle chromatism, using tabulated values for the zero-angle given in the FORS2 user manual \footnote{\url{http://www.eso.org/sci/facilities/paranal/instruments/fors/doc/VLT-MAN-ESO-13100-1543_P07.pdf}}. The wavelength scale was corrected to the rest frame using the host-galaxy redshift. To derive the continuum polarization from the spectropolarimetric data, we adopted the wavelength ranges between 6800 and 7200 {\AA} and between 7820 and 8140 {\AA}, as in \citet[][]{Nagao2024a}.

In the case of imaging polarimetry, the same instrumental set-up was adopted, using two broad-band filters ($V$ and $R$) for NOT, instead of the grism in the optical path. We applied bias subtraction and flat-field correction to the object frames and then performed aperture photometry on ordinary and extraordinary sources of the object. For the aperture photometry, we adopted the same procedures as in \citet[][]{Nagao2024b}, using an aperture size that is twice as large as the full-width-half-maximum (FWHM) of the ordinary beam’s point-spread function and a background region whose inner and outer radii are twice and four times as large as the FWHM, respectively. Based on the derived values, we calculated the polarization degree and angle. When calculating the polarization degrees, we subtracted the polarization bias, following \citet[][]{Wang1997}.

\subsection{Spectroscopy}

We obtained optical spectra of SNe~2018khh, 2021adxl, 2022oo, and 2022mma, using ALFOSC/NOT (with grism \#4, giving a wavelength coverage of 3200–9600 {\AA} and a spectral resolution of $\sim 360$), FORS2/VLT (with grism 300V) and the Goodman High Throughput Spectrograph (Goodman HTS) mounted on the Southern Astrophysical Research (SOAR) telescope. The FORS2/VLT spectrum of SN 2018khh was obtained as part of the FOSSIL survey (see \citealt{Kuncarayakti2022}). We also conducted spectroscopic observations for SNe~2018khh and 2022oo using the ESO Faint Object Spectrograph and Camera version 2 (EFOSC2; with grism \#13) mounted on the New Technology Telescope (NTT) at La Silla Observatory in Chile as part of the extended Public ESO Survey for Transient Objects \citep[ePESSTO+;][]{Smartt2015}. The log of spectroscopic observations are provided in Tables~\ref{tab:spec_21adxl}-\ref{tab:spec_18khh}. 

The ALFOSC/NOT spectra were reduced using the alfoscgui pipeline\footnote{FOSCGUI is a graphical user interface aimed at extracting SN spectroscopy and photometry obtained with FOSC-like instruments. It was developed by E. Cappellaro. A package description can be found at \url{https://sngroup.oapd.inaf.it/foscgui.html}}. This pipeline includes the following procedures: overscan, bias, and flat-field corrections, cosmic-ray removal, extraction of a one-dimensional spectrum, and sky subtraction. The wavelength calibration was performed by comparison with arc lamps. The flux scale of the extracted spectra were calibrated using a sensitivity function derived from a standard star observed on the same night. We reduced the EFOSC2/NTT and FORS2/VLT data using the PESSTO\footnote{https://github.com/svalenti/pessto} and ESOReflex \citep[][]{Freudling2013} pipelines, respectively, which include standard tasks such as correction for bias, flat field, and wavelength scale using arc lamp. The flux calibration was conduced based on observations of a standard star. The Goodman spectra of SN 2018khh were observed at the parallactic angle using the $1\farcs0$ slit with the 400 lines/mm grism providing a spectral resolution of $R\sim1000$. The spectra were reduced, wavelength and flux calibrated using standard procedures in IRAF \citep[see][]{Cartier2024}. The wavelength calibration was checked against sky lines, and the telluric correction was performed using a standard star observed on the same night. Most of the Goodman spectra of SN 2018khh were presented in \citet[][]{Jacobson-Galan2024}.

\section{Results} \label{sec:result}

\subsection{Photometric properties} \label{sec:photo_prop}

   \begin{figure*}
   \centering
            \includegraphics[width=\hsize]{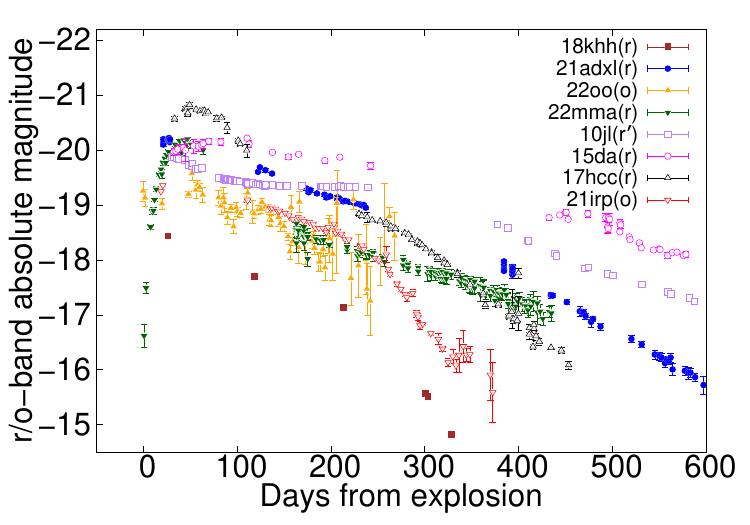}
      \caption{Light curves of the SNe in our sample. All data are corrected for the assumed extinction (see Section~\ref{sec:sn_sample}). The data points from the same night are binned. The data for SNe~2010jl, 2015da, 2017hcc and 2021irp are taken from \citet[][]{Fransson2014}, \citet[][]{Tartaglia2020}, \citet[][]{Moran2023}, and \citet[][]{Reynolds2025a}, respectively.      
              }
      \label{fig:LC}
   \end{figure*}

Figure~\ref{fig:LC} shows the $r$/$o$-band LCs of the SNe in our sample. These SNe are intrinsically bright and long-lived, showing absolute magnitudes between $\sim -17$ and $\sim -20$ mag even at $\sim 200$ days after the explosion. In contrast, Type~IIP SNe show absolute magnitudes between $\sim -12$ and $\sim -15$ mag in the $V$ band already at the beginning of the tail phase, i.e., $\sim 100$ days after the explosion \citep[][]{Anderson2014}. The LC of SNe~2010jl, 2017hcc and 2022mma have a bumpy structure during the first $\sim$50--200 days after the explosion followed by a relatively linear decline, while those of SNe~2015da, 2021adxl and 2022oo show a relatively linear decline from early phases. Due to lack of data, the early-phase evolution of SNe~2018khh and 2021irp is unclear.

The peak absolute magnitudes in the $r$/$o$/$g$ bands spreads over a wide range from $\sim-18.7$ to $\sim-21$ mag. Here, since our sparse data do not cover the LC peaks of SNe~2018khh, 2021irp and 2021adxl, their actual peak magnitudes might be brighter than the observed maxima. 
For SN~2022mma, it takes a long time to reach the LC peak ($\sim 50$ days), which is much longer than for Type~IIP/L SNe \citep[e.g., $\sim 7.5$ days in the $g$ band;][]{Gonzalez-Gaitan2015}. The rise time of SN~2021irp is also estimated to be long \citep[$\gtrsim 20.5\pm 3$ days;][]{Reynolds2025a}, although the precise value is unclear due to the lack of observational coverage around the LC peak. SNe~2015da and 2017hcc also have long rise times of $100 \pm 3$ days in the $R$ band \citep[][]{Tartaglia2020} and $57 \pm 2$ days in the ATLAS $o$ band \citep[][]{Moran2023}, respectively. On the other hand, the rise time for SN~2018khh is short. The first detection, which was $\sim 3$ days after the last non-detection \citep[][]{Brimacombe2018}, was brightest in all our observation epochs (see Figure~\ref{fig:LC_18khh}). Thus, the brightness peak should be at a timing between the last non-detection and the second detection, and the rise time should be $\lesssim 28$ days and probably much shorter than this. The rise times for SNe~2010jl, 2021adxl and 2022oo are unclear due to the lack of a good estimation of the explosion dates and/or good observational coverage of the LC peaks.

The decline rates of the LCs of the SNe in our sample are all similarly slow, although this is one of our criteria for the target selection. In the case of SN~2021irp, the decline rate in the $o$-band LC is 0.0076 mag day$^{-1}$ from 132 to 230 days, and 0.027 mag day$^{-1}$ between 250 and 300 days, although the latter accelerated decline is interpreted not as an intrinsic feature but as the apparent effect of the attenuation by newly formed dust \citep[][]{Reynolds2025a}. The decline rates are from $\sim 0.001$ to $\sim 0.007$ mag day$^{-1}$ for SNe~2010jl (in the $r^{\prime}$ band, from $\sim 100$ to $\sim 250$ days), 2015da (in the $r$ band, from $\sim 100$ to $\sim 600$ days), 2017hcc (in the $r$ band, from $\sim 230$ to $\sim 280$ days), 2018khh (in the $r$ band, from $\sim 100$ to $\sim 200$ days), 2021adxl (in the $r$ band, from $\sim 200$ to $\sim 400$ days), 2022oo (in the $o$ band, from $\sim 50$ to $\sim 150$ days), and 2022mma (in the $r$ band, from $\sim 200$ to $\sim 400$ days). All these decline rates are smaller than the value of the $^{56}$Co decay with full gamma-ray trapping, i.e., 0.0098 mag day$^{-1}$.

SNe~2010jl, 2017hcc, 2018khh, and 2021adxl seem to have some kind of acceleration of the LC decline at certain points between $\sim 250$ and $\sim350$ days, around $\sim 280$ days, between $\sim 200$ and $\sim250$ days, and between $\sim 250$ and $\sim350$ days after the explosion, respectively. On the other hand, SNe~2015da and 2022mma do not show clear signs of an accelerated decline at least until $\sim 900$ days \citep[][]{Tartaglia2020} and $\sim 400$ days after the explosion, respectively. SN~2015da even shows deceleration of the decline at $\sim 900$ days after the explosion \citep[][]{Tartaglia2020}.

\subsection{Spectroscopic properties} \label{sec:spec_prop}

\begin{figure*}
            \includegraphics[width=0.5\hsize]{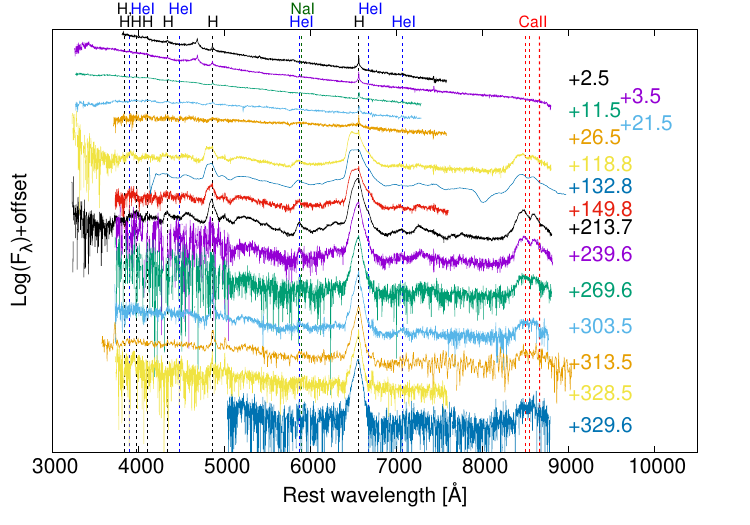}
            \includegraphics[width=0.5\hsize]{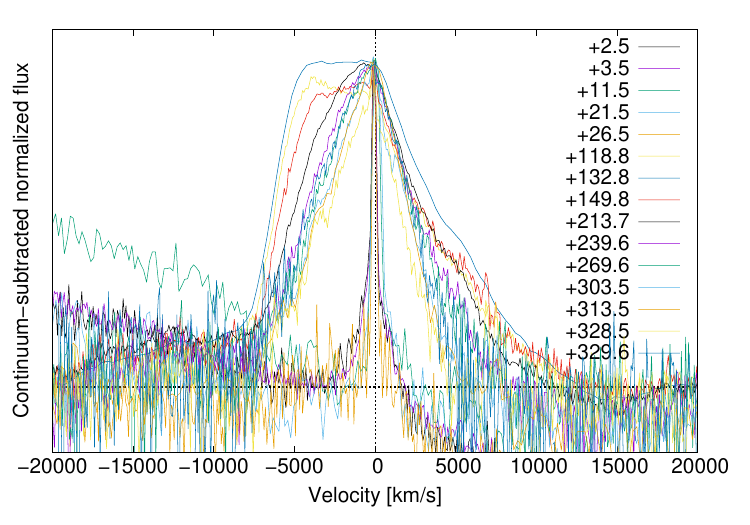}
      \caption{Spectra of SN~2018khh. Left panel: Spectral evolution. All spectra are corrected for the assumed extinction (see Section~\ref{sec:sn_sample}). Right panel: Time evolution of the H$\alpha$ line. The continuum level is assumed to be a constant value and estimated from the averaged value from $-3500$ to $-2500$ km s$^{-1}$ and around $-20000$ km s$^{-1}$ in the velocity space, for the first 5 epochs and the later epochs, respectively. The vertical and horizontal dotted lines show the zero velocity and the assumed continuum level, respectively.
              }
      \label{fig:fig2}
   \end{figure*}

\begin{figure*}
            \includegraphics[width=0.5\hsize]{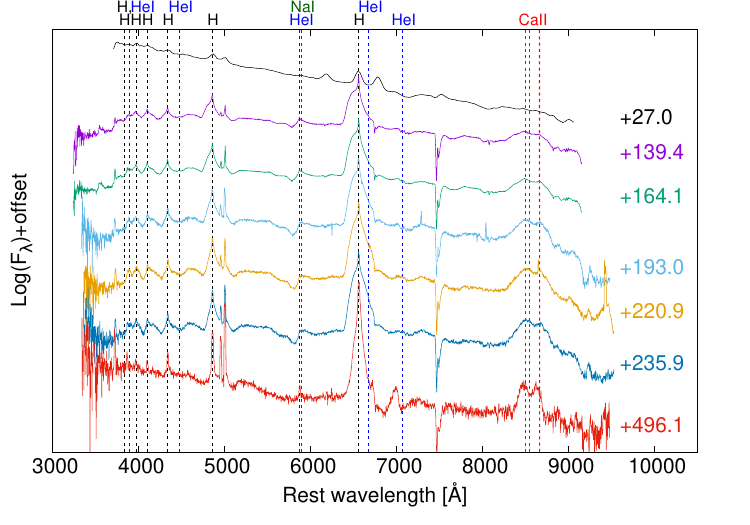}
            \includegraphics[width=0.5\hsize]{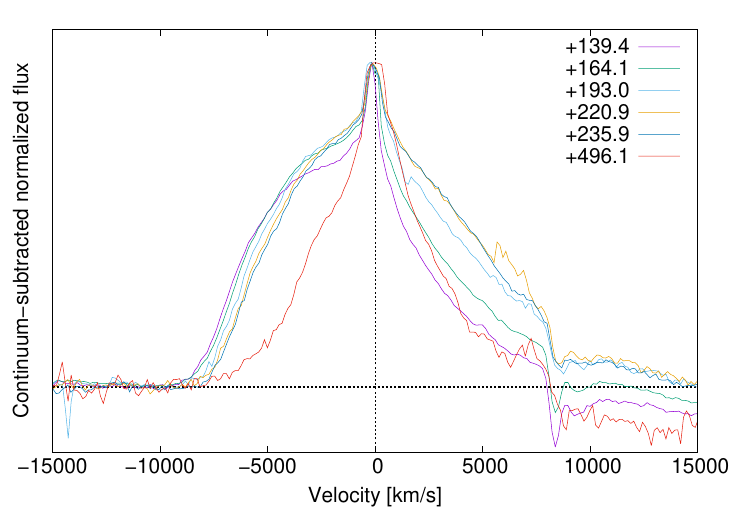}
      \caption{Same as Fig.~\ref{fig:fig2} but for SN~2021adxl. The continuum level is assumed to be a constant value and estimated from the averaged value from $-15000$ to $-10000$ km s$^{-1}$ in the velocity space. 
              }
      \label{fig:fig3}
   \end{figure*}

\begin{figure*}
            \includegraphics[width=0.5\hsize]{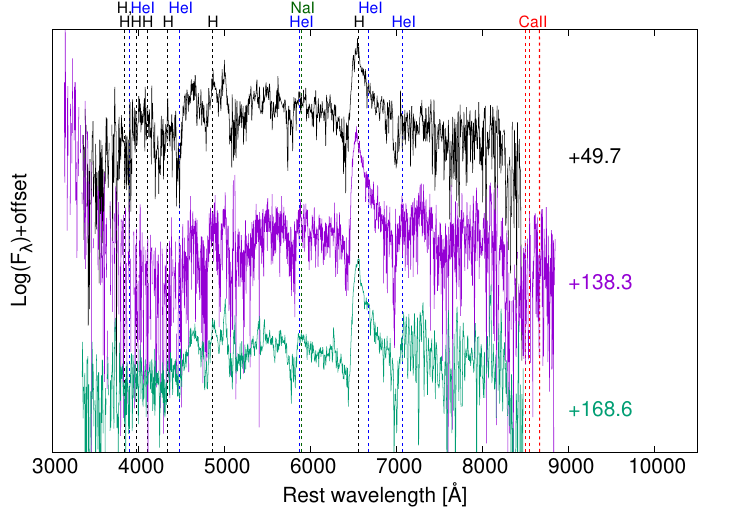}
            \includegraphics[width=0.5\hsize]{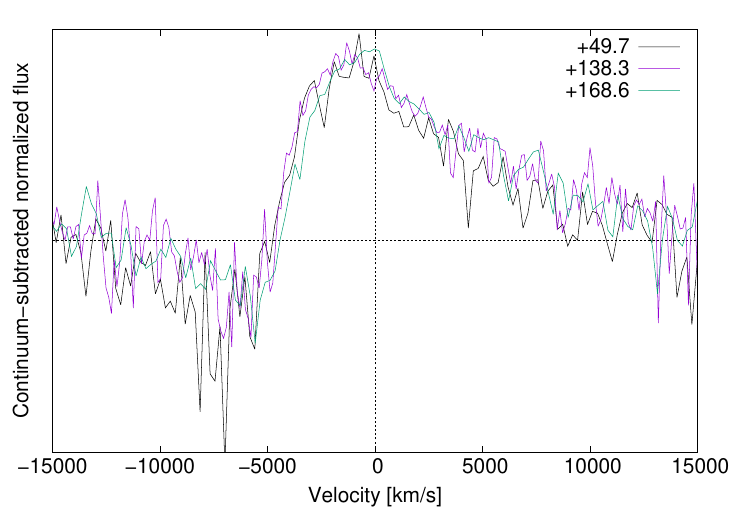}
      \caption{Same as Fig.~\ref{fig:fig2} but for SN~2022oo. The continuum level is assumed to be a constant value and estimated from the averaged value from $-15000$ to $-14000$ km s$^{-1}$ in the velocity space.
              }
      \label{fig:fig4}
   \end{figure*}

\begin{figure*}
            \includegraphics[width=0.5\hsize]{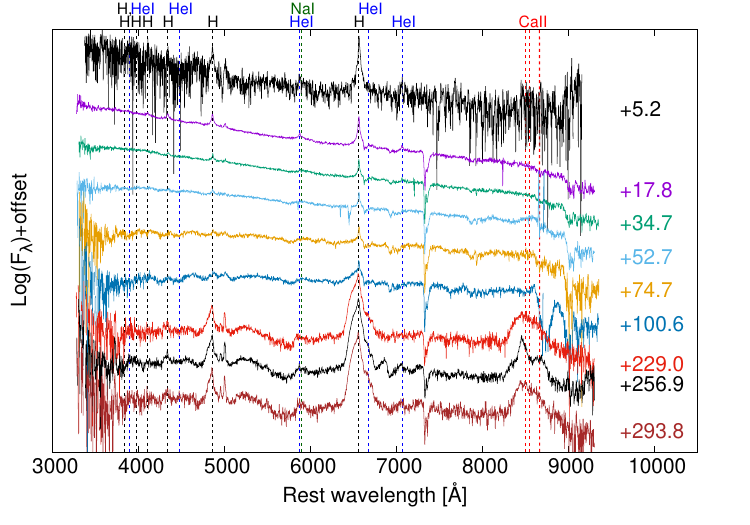}
            \includegraphics[width=0.5\hsize]{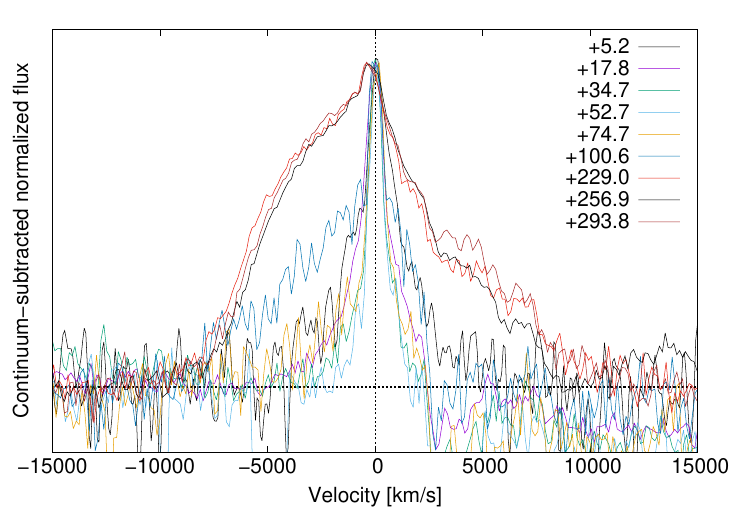}
      \caption{Same as Fig.~\ref{fig:fig2} but for SN~2022mma. The continuum level is assumed to be a constant value and estimated from the averaged value from $-15000$ to $-10000$ km s$^{-1}$ in the velocity space.
              }
      \label{fig:fig5}
   \end{figure*}

\begin{figure*}
   \centering
            \includegraphics[width=\hsize]{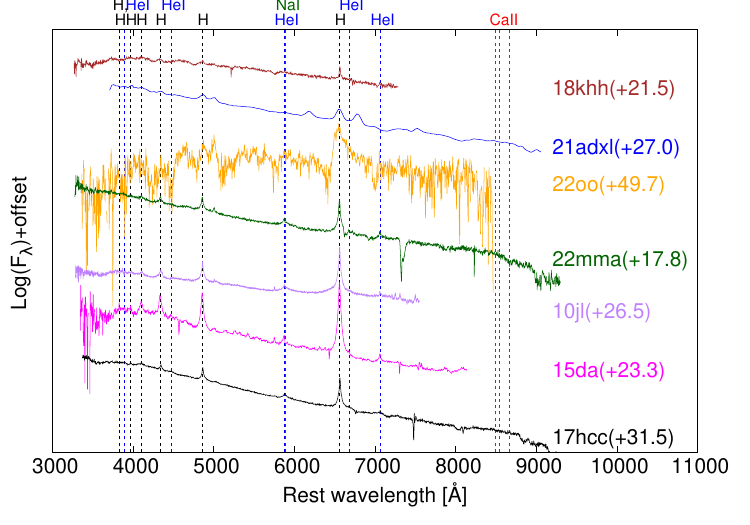}
      \caption{Comparison of the early-phase spectra of the SNe in our sample. All spectra are corrected for extinction (see Section~\ref{sec:sn_sample}). The data for SNe~2010jl, 2015da, 2017hcc and 2021adxl are taken from \citet[][]{Smith2012}, \citet[][]{Tartaglia2020}, \citet[][]{Moran2023} and \citet[][]{Brennan2024}, respectively.
      }
      \label{fig:fig6}
   \end{figure*}

\begin{figure*}
   \centering
            \includegraphics[width=\hsize]{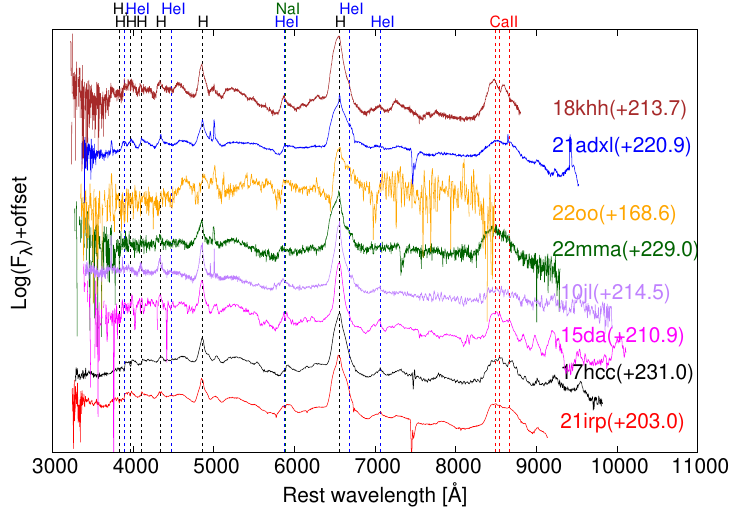}
      \caption{Same as Figure~\ref{fig:fig6}, but for the late-phase spectra.
The data for SNe~2010jl, 2015da, 2017hcc and 2021irp are taken from \citet[][]{Smith2012}, \citet[][]{Tartaglia2020}, \citet[][]{Moran2023} and \citet[][]{Reynolds2025a}.
      }
      \label{fig:fig7}
   \end{figure*}

Figures~\ref{fig:fig2}-\ref{fig:fig5} show the evolution of the spectra of the SNe in our sample. The comparison of the early-phase and late-phase spectra between the SNe are shown in Figures~\ref{fig:fig6} and \ref{fig:fig7}, respectively. They show photospheric spectra with a continuum radiation and emission lines from allowed lines (e.g., Balmer lines, He~I $\lambda$5876, Na~I~D $\lambda\lambda$ 5890, 5896, He~I $\lambda$7065, or Ca~II NIR triplet $\lambda\lambda\lambda$ 8498, 8542, 8662) for several hundreds of days.

The early spectra show narrow ionized/excited lines with Lorentzian wings (so-called flash-ionization lines; e.g., Balmer lines, He~I and/or He~II lines) superposed on a blue continuum in SNe~2018khh, 2021adxl and 2022mma. These features are also observed in early phases of SNe interacting with CSM, for example, in Type~IIP and IIn SNe, and are regarded as a sign of a strong CSM interaction. As these narrow lines become weaker, the continuum becomes redder and broad emission lines emerge around $\sim 50-100$ days after the explosion in SNe~2018khh, 2021adxl and 2022mma, as in SN~2021irp. These features are similar to those of some Type IIL SNe \citep[e.g.,][]{Faran2014}, some broad-line-dominated Type~II superluminous SNe \citep[e.g.,][]{Kangas2022} and some luminous Type~II SNe \citep[e.g.,][]{Pessi2023}. At the same time, the so-called Fe bump (an excess flux at $\lambda \lesssim 5500$ \AA, which is interpreted to be due to ionized/excited Fe lines) develops. Unlike the other SNe, SN~2022oo shows a significant absorption part in the similarly broad Balmer emission lines at least during the period between $\sim 50-170$ days. The spectroscopic features of the narrow ionized/excited lines in the early spectra and the Fe bump in the late spectra are often seen in other interacting SNe, e.g., in Type IIn, Ia-CSM, interacting Ibc SNe, Ibn, or Icn SNe \citep[e.g.,][]{Gangopadhyay2025,Sharma2023,Kuncarayakti2018,Pastorello2008,Pellegrino2022b}. The main spectral features such as the continuum and broad Balmer lines are similar to those of Type II SNe, in particular Type~IIL SNe, which typically have Balmer lines with shallower absorption parts than Type~IIP SNe. This suggests that these SNe have a photosphere in the SN ejecta not for $\sim 100$ days as in prototypical Type~II SNe but at least for several hundreds of days. In these SNe we have not observed nebular-phase spectra, dominated by forbidden lines suggesting radiation from low-density ionized gas in the SN ejecta, even at the latest epochs of our observations (e.g., $\sim 500$ days in the case of SN~2021adxl) and $\sim 1500$ days in the case of SN~2015da \citep[][]{Tartaglia2020}. This lack of nebular features is also seen in long-lived Type~IIn SNe at these phases \citep[e.g.,][]{Zhang2012, Fox2013}.

Our SNe show diverse shapes of the broad Balmer lines. In the case of SN~2021irp \citep[see][]{Reynolds2025a}, the H$\alpha$ line consists of a narrow component with FWHM of 2100 km s$^{-1}$ and a broad component with 7700 km s$^{-1}$ at $\sim 200$ days after the explosion. The peaks of these components are blueshifted from the rest wavelength of the H$\alpha$ line. This line shape remains for a while, and its red part is slightly decreased at $\sim 300$ days. At $\sim 350$ days, the red part of the line profile is further reduced, and a slight erosion of the bluer part is observed as well. Eventually at $\sim 530$ days, the blue part also becomes narrower. The line-shape evolution before $\sim200$ days is unclear, due to the lack of observations.

SN~2021adxl also has an H$\alpha$ line with narrow and broad components. During the phases from $\sim 140$ to $\sim 240$ days, the width of the blue part of the broad component is relatively constant at $\sim 4000-5000$ km s$^{-1}$, while that of the red part increases from $\sim 2500$ km s$^{-1}$ to $\sim 4000$ km s$^{-1}$. Then, at $\sim 500$ days, both the blue and red parts of the broad component decrease to $\sim 2500$ km s$^{-1}$. The narrow component is relatively stable at $\sim 1000$ km s$^{-1}$ over the whole period of time with its width slightly increasing. SN~2022mma shows an H$\alpha$ line with a narrow component dominating with possible Lorenzian wings ($\sim 1000$ km s$^{-1}$) until $\sim 75$ days. A broad component emerges at $\sim 100$ days and becomes dominant by $\sim 230$ days. After that, the shape does not evolve until $\sim 300$ days. The width of the blue part of the broad component is larger than that of the red part throughout the evolution. SN~2022oo shows a slightly different evolution for the H$\alpha$ line. From $\sim 50$ to $\sim 170$ days, it exhibits a broad P-Cygni shape with a significant absorption part. The absorption minimum is at $\sim 8000$ km s$^{-1}$ at $\sim 50$ days, and at $\sim 6000$ km s$^{-1}$ around $\sim 150$ days. The H$\alpha$ line of SN~2018khh also follows a similar evolution. At early phases, it shows a narrow component with Lorentzian wings. At $\sim 120$ days, it develops a broad H$\alpha$ line with a boxy shape. As time goes, the line becomes narrower.

\begin{figure}
   \centering
            \includegraphics[width=\hsize]{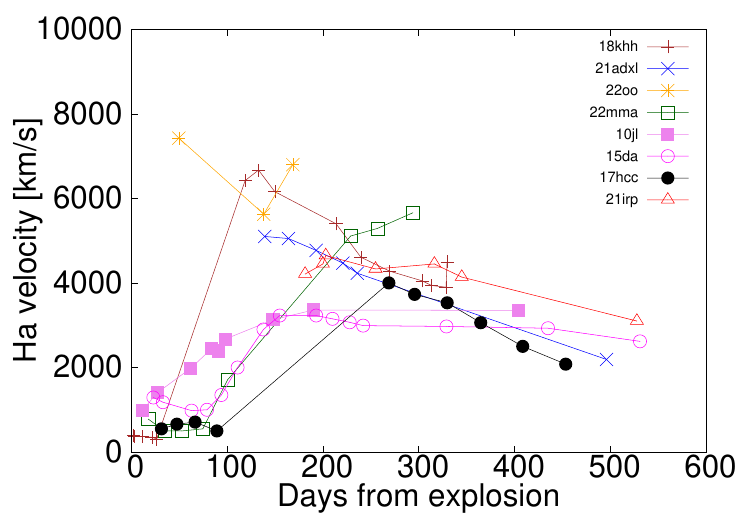}
      \caption{Evolution of the H$\alpha$ velocity. The velocity measurements for SNe~2010jl, 2015da, 2017hcc and 2021irp are conducted using selected low-resolution spectra from \citet[][]{Smith2012,Jencson2016}, \citet[][]{Tartaglia2020}, \citet[][]{Moran2023} and \citet[][]{Reynolds2025a}.
      }
      \label{fig:fig8}
   \end{figure}

The overall behavior of the H$\alpha$ line of these SNe resembles those of the well-observed similar SNe~2010jl \citep[see Figure~12 in][]{Fransson2014}, 2015da \citep[see Figures~11 and 13 in][]{Tartaglia2020}, 2017hcc \citep[see Figure~11 in][]{Moran2023}, and 2021irp \citep[see Figure~10 in][]{Reynolds2025a}. At early phases (until $\sim 100$ days since the explosion), the narrow component with Lorentzian wings dominates. Then, the contribution from the broad component starts to increase as shown by the increase of the line width. For a while, the broad component remains with the same width. Sometimes the redder parts are dramatically decreased. Finally (from $\sim 300$ days since the explosion) the line becomes narrower with time.
For quantifying the time evolution of the width of the H$\alpha$ line, we derive the velocity from the width of its blue part at the half maximum, except for SN~2022oo. As for SN~2022oo, we measure the H$\alpha$ velocity from the absorption minimum of its P-Cygni profile. The estimated velocities are shown in Figure~\ref{fig:fig8}.

\subsection{Polarimetric properties} \label{sec:pol_prop}

\begin{figure*}
   \centering
            \includegraphics[width=\hsize]{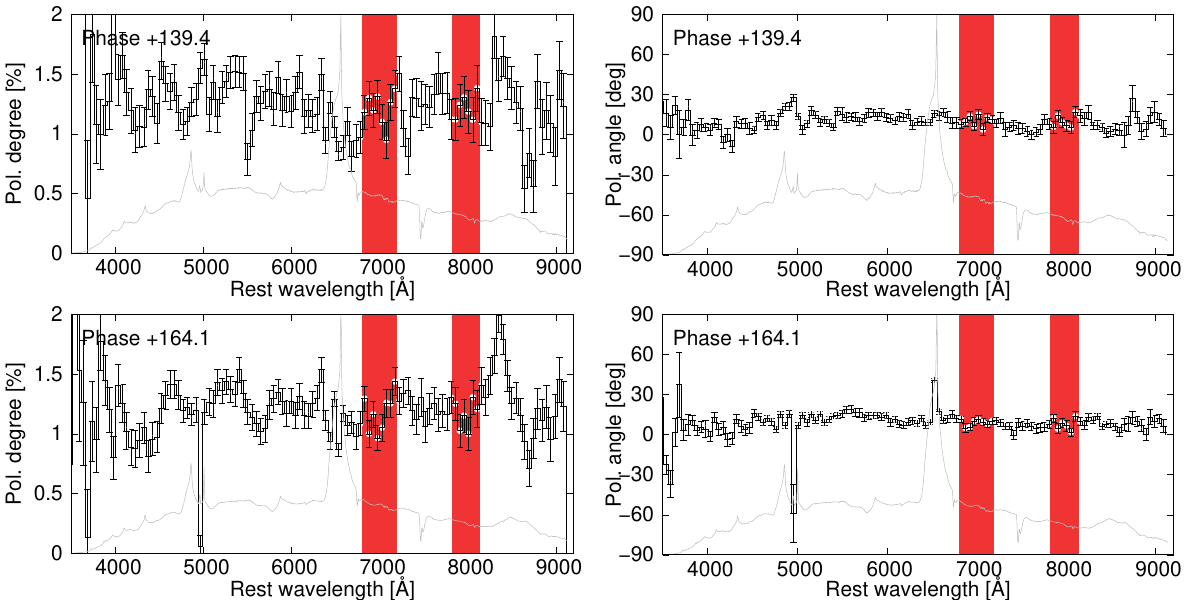}
      \caption{Polarization degrees and angles of SN~2021adxl at Phases +139.35 and +164.07 days. The red hatching indicates the adopted wavelength regions for the continuum polarization estimate.
      }
      \label{fig:fig9}
   \end{figure*}

Figure~\ref{fig:fig9} shows the polarization spectra of SN~2021adxl at +139.4 and 164.1 days. The spectra show continuum polarization with high degrees ($\sim 1.2$ \%) and relatively constant angles ($\sim 10$ degrees), as well as some line polarization/depolarization in the emission lines. In general, this continuum polarization observed in SNe can originate not only from the SN intrinsic polarization but also from the interstellar polarization (ISP). However, the empirical relation between the extinction and the consequent polarization by dust \citep[$P \lesssim 9 E(B-V)$;][]{Serkowski1975} indicates that its ISP should likely be smaller than $\sim 0.2$ \% for the assumed extinction ($E(B-V)=0.026$ mag). Since the observed polarization degrees are much higher than this maximum ISP value, we simply ignore the ISP and assume that the observed polarization is purely the intrinsic SN polarization. 

\begin{figure*}
   \centering
            \includegraphics[width=0.49\hsize]{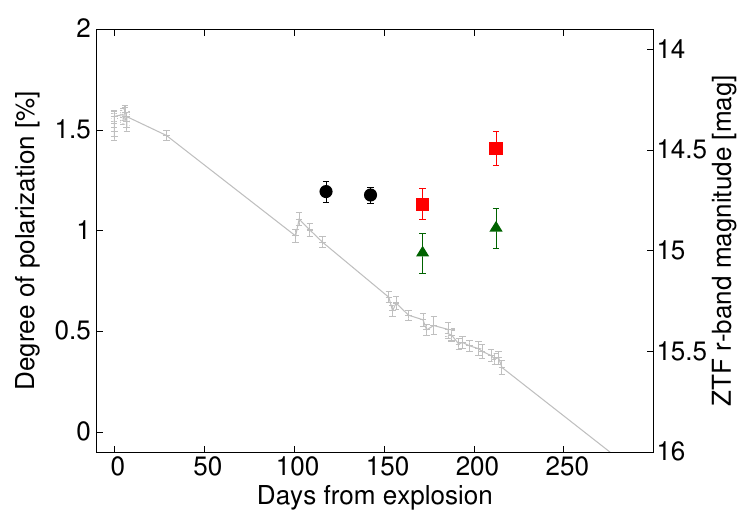}
            \includegraphics[width=0.49\hsize]{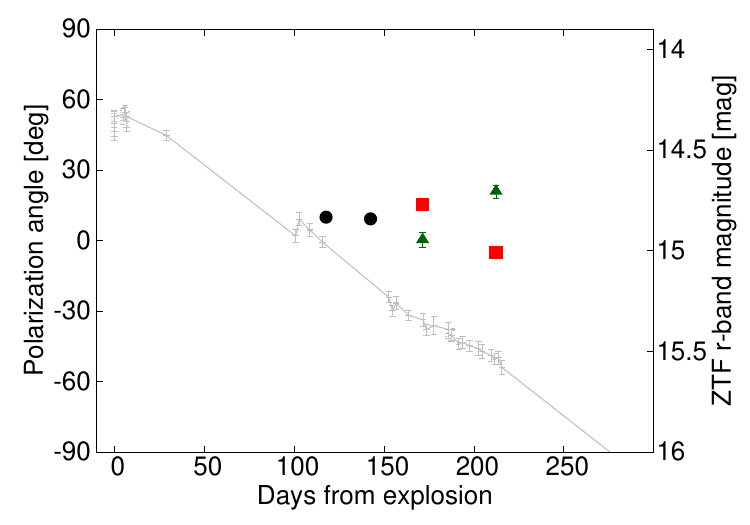}
      \caption{Polarization degrees and angles of the continuum (black), $V$-band (green) and $R$-band (red) polarization of SN~2021adxl. The gray points connected with a line trace the r-band light curve of SN~2021adxl.}
      \label{fig:fig10}
   \end{figure*}

We estimate the degree and angle of the continuum polarization at each epoch by averaging the signals of the polarization spectrum in the wavelength regions from 6800 {\AA} to 7200 {\AA} and from 7820 {\AA} to 8140 {\AA} \citep[see the red hatching in Figure~\ref{fig:fig9}; e.g.,][]{Chornock2010, Nagao2019, Nagao2021, Nagao2024a, Reynolds2025b}. The time evolution of the continuum polarization and the $V$- and $R$-band polarization is shown in Figure~\ref{fig:fig10}. Although the broad-band polarization can generally be affected by the line polarization/depolarization, all the data points show relatively constant degrees and angles over the whole period of the observations (from $\sim 140$ to $\sim 230$ days). This suggests that this SN has an aspherical photosphere with a relatively constant orientation of the axis with time.

\begin{figure*}
   \centering
            \includegraphics[width=0.49\hsize]{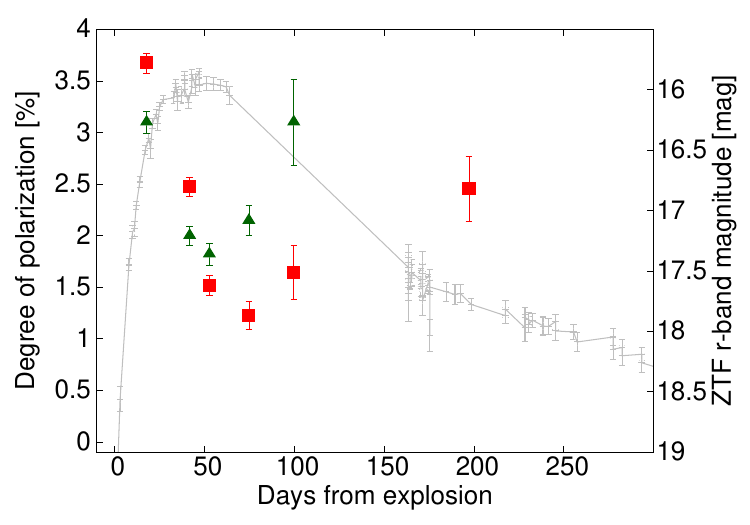}
            \includegraphics[width=0.49\hsize]{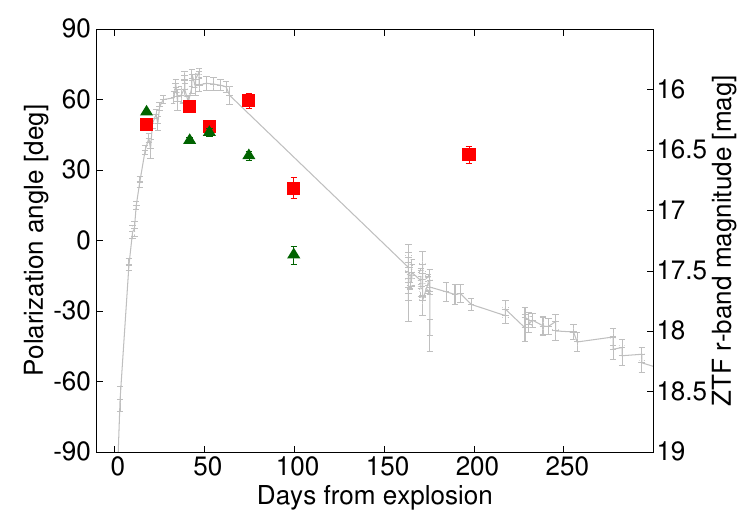}
      \caption{Same as Fig.~\ref{fig:fig11}, but for SN~2022mma.}
      \label{fig:fig11}
   \end{figure*}

Figure~\ref{fig:fig11} shows the time evolution of the $V$- and $R$-band polarization in SN~2022mma. Both the $V$- and $R$-band data show very high polarization degrees ($\sim 3.5$\%) with a relatively constant polarization angle ($\sim 50$ degrees) at early phases before the LC peak, and a decline to $\sim 1.5$\% at $\sim 70$ days after the explosion. After that, they exhibit a slight increase of the polarization degrees and a slight change of the polarization angles, as well as a light deviation between them. This later-epoch behavior would be due to the contamination of some components of line polarization/depolarization, creating different values for the $V$- and $R$-band polarization, and could possibly also be due to other components of continuum polarization. As in the case of SN~2021adxl, the empirical relation between the extinction and the polarization by dust also indicates minimal ISP for SN~2022mma ($P_{\rm{ISP}} \lesssim 0.2$ \% for the assumed extinction of $E(B-V)=0.023$ mag). Therefore, the observed early-phase polarization should originate from the SN itself and suggests a very aspherical shape of the photosphere.

\begin{figure*}
   \centering
            \includegraphics[width=\hsize]{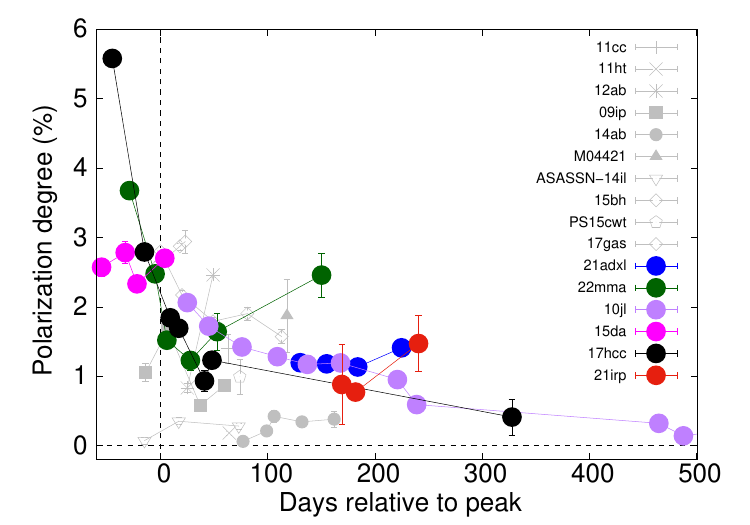}
      \caption{Time evolution of the continuum polarization of the SNe in our sample with polarimetric data (colored points), compared with those of Type~IIn SNe presented in \citet[][]{Bilinski2024} (gray points). The $R$-band polarization is regarded as the continuum polarization for SNe~2021adxl and 2022mma. The data for SNe~2010jl, 2015da, 2017hcc, and 2021irp are taken from \citet[][and references therein]{Bilinski2024} and from \citet[][]{Reynolds2025b}.
      }
      \label{fig:fig12}
   \end{figure*}

Figure~\ref{fig:fig12} shows the time evolution of the intrinsic SN polarization in our sample, compared with those of SN~2021irp and other Type~IIn SNe. The polarization of most Type~IIn and 21irp-like SNe shows very high values ($\gtrsim 3$\%) at early phases, rapid declines to $\sim 1$\% at several tens of days after the LC peak, and then slower declines for another several hundreds of days, except for a few Type~IIn SNe. All the 21irp-like SNe in this paper seem to follow this general trend, suggesting the presence of some common CSM geometries for Type~II SNe interacting with hydrogen-rich CSM.

\section{Discussion} \label{sec:discussion}

\subsection{The dominant energy source in 21irp-like SNe}  \label{sec:discussion_energy_source}

The SNe in our sample are more luminous and of longer duration than the most common subtype of Type~II SNe, that is, Type IIP SNe (see Section~\ref{sec:photo_prop}). This requires an additional energy source rather than the usual ones in Type~IIP SNe, i.e., the thermal energy deposited by the explosion shock and the decay energy of radioactive elements such as $^{56}$Ni/$^{56}$Co. They show quite common photometric, spectroscopic and polarimetric properties: they are luminous and long-lived, and show photospheric spectra (a continuum plus allowed lines) with broad Balmer lines and high polarization ($\sim$ a few \%). As discussed in the case of SN~2021irp in \citet[][]{Reynolds2025a}, the photometric, spectroscopic and polarimetric properties of these SNe do not support a large amount of $^{56}$Ni or the presence of a central engine (e.g., magnetar, fallback accretion). In particular, their LC decline rates at late phases do not match with those predicted for $^{56}$Co decay, magnetar power \citep[$t^{-2}$, e.g.,][]{Dessart2018} or fallback accretion \citep[$t^{-5/3}$, e.g.,][]{Dexter2013}. In addition, the line shapes especially of the Balmer lines, where the emission part is dominant compared to the absorption part, do not support scenarios with the ejecta heated from an energy source in the central parts of the ejecta.

Another important argument for excluding the Ni/Co and central engine scenarios comes from the fact that the shape of the photosphere of 21irp-like SNe should be aspherical. The locations of the emitting regions of these SNe, estimated from the line velocities, are much larger than the blackbody radius. Following the discussion in Section~6.1 in \citet[][]{Reynolds2025a}, the location of the emitting region at 300 days should be $r\gtrsim 2500$ km s$^{-1}$ $\times$ 300 days $\sim 6 \times 10^{15}$ cm, while the blackbody radius is $r\lesssim 2\times 10^{15}$ cm estimated from the brightness and temperature of these SNe (see Sections~\ref{sec:photo_prop} and \ref{sec:spec_prop}). This argument is supported by the high polarization degrees for the SNe with the polarimetric observations (see Section~\ref{sec:pol_prop}). Therefore, as in the case of SN~2021irp, we conclude the CSM interaction to be the main energy source for the radiation of these SNe, and that the CSM interaction should be aspherical to some extent. 

Many of these SNe show narrow components of Balmer lines at least at early phases (SNe~2010jl, 2015da, 2017hcc, 2018khh, 2021adxl, and 2022mma; see Figure~\ref{fig:fig6}) and were indeed classified as Type~IIn SNe, indicating the presence of CSM interaction. At the same time, several SNe (SNe~2010jl, 2015da, 2017hcc, 2018khh, 2021irp, 2021adxl, and 2022mma; see Figure~\ref{fig:fig7}) have a so-called ``Fe bump" in their late-phase spectra, which is often seen in SNe interacting with CSM. Most importantly, all the observational properties of these SNe can be at least qualitatively explained by the aspherical CSM interaction scenario. The general picture for the aspherical CSM interaction has been described in \citet[][; see their Figure~6]{Reynolds2025b}. Here, parts of the SN ejecta collide with aspherical CSM, creating shocked regions. This CSM interaction creates a luminous and long-lived LC as well as narrow Balmer lines at early phases. The other parts of the ejecta in the CSM-free directions can freely expand, and cover the shock regions at some point of time. Optically-thick regions, i.e., the photospheres, can be created around the shocked regions in the SN ejecta, which is the origin of the long-lived broad-line-dominated photospheric spectra in 21irp-like SNe. These aspherical photospheres introduce the high continuum polarization degrees observed in 21irp-like SNe.

\subsection{CSM properties} \label{sec:discussion_csm_properties}

In this subsection, we discuss the CSM properties of the SNe in our sample. In the above subsection (Section~\ref{sec:discussion_energy_source}), we conclude that the CSM interaction around the 21irp-like SNe should be aspherical based on the relation between the location of the emitting region and the blackbody radius and on their high polarization (see Section~\ref{sec:pol_prop}). The next question we consider is: which aspherical CSM structures are consistent with our observations of 21irp-like SNe? All the SNe in our sample have relatively smooth LC shapes (see Figure~\ref{fig:LC}), although this is unclear in the case of SN~2022oo due to the poor quality of the photometry. In addition, the polarization angles of SNe~2021irp \citep[][]{Reynolds2025b}, 2021adxl and 2022mma are relatively constant with time. We note that, the late-phase evolution for SN~2022mma does not follow this trend, but this might be due to the possible contamination of the line polarization, implied by the discrepancy of the $V$- and $R$-band polarization. These observational features support some continuous distribution in the radial direction, excluding randomly distributed multiple small CSM clumps. Even if they have such clumpy CSM distributions, the interaction with one major CSM clump should last at least for several hundreds of days. As discussed in \citet[][]{Reynolds2025b}, the H$\alpha$ line profiles in these SNe smoothly spread both to blue and red directions from the rest wavelength (see Section~\ref{sec:app_obs}). This requires the emitting regions to smoothly spread between the approaching and receding sides in the line of sight to the observer. This is consistent with a disk-like distribution of CSM as proposed for SN~2021irp in \citet[][]{Reynolds2025b}. In this interpretation, some red parts of the H$\alpha$ line can be hidden by the CMS disk and/or the photosphere in the near side, depending on the viewing angle \citep[see Figure~6][]{Reynolds2025b}. Indeed, most 21irp-like SNe have, more or less, some reduction of the red parts of the H$\alpha$ lines compared to the blue parts, that is, blue-shifted line profiles.

The general observational properties of the 21irp-like SNe are similar: luminous and long-lived LCs, long-lived broad-line-dominated photospheric spectra, and high polarization. However, the details of these observational properties are different from each other, as we see in Section~\ref{sec:result}. This could reflect the different CSM properties, different SN ejecta properties and/or different viewing angles.
The diverse properties of the LCs (see Section~4.1) most likely mainly represent the different distributions of the CSM and/or the different SN properties, although some viewing angle effects could also exist as suggested in, e.g., \citet[][]{Suzuki2019}. For SN~2021irp, \citet[][]{Reynolds2025b} determined the CSM properties as a disk-like CSM with a corresponding mass and half-opening angle $\gtrsim 2$ M$_{\odot}$ of and $30-50$ degrees, respectively, based on the bolometric light curve evolution and the high polarization. Since the difference of the optical brightness between the other SNe in our sample and SN~2021irp is roughly within $\pm 2$ mag (i.e., the flux difference within a factor of $\sim 6$; see Figure~\ref{fig:LC}), the overall difference of the brightness reflects different scales of the CSM, i.e., the total CSM masses ranging from $\sim 0.1$ to a few tens of M$_{\odot}$. At the same time, the LCs show slightly different evolutions with different decline rates. For example, the LCs of SNe~2010jl, 2015da, 2021adxl, and 2022mma have similar peak absolute magnitudes, while the late-phase absolute magnitudes (at $\sim 200-400$ days) have a larger discrepancy with $\sim 2$ mag. This behavior cannot be explained only by the different scales of the CSM density but requires different radial distributions and/or different opening angles of the CSM disk. 

The viewing angle can also play an important role in the LC shapes of these SNe. The optical-depth effects by the SN ejecta depend on the viewing angle. At the same time, since the photospheres of these SNe are aspherical, a different brightness is expected for different viewing angles even towards a system with identical SN and CSM parameters. For evaluating this effect, we would need to conduct radiation hydrodynamic simulations as in \citet[][]{Suzuki2019} but with a proper treatment for the ionization/recombination of hydrogen. Moreover, newly formed dust in the SN ejecta and/or the interaction shocks can change the LC shapes in these systems. Indeed, \citet[][]{Reynolds2025a} pointed out that the acceleration of the LC decline in SN~2021irp is due to the effects by the newly formed dust. SN~2018khh has an earlier timing of such acceleration of the LC decline (around $\sim$180-250 days), while SNe~2010jl, 2017hcc and 2021adxl seem to have a later timing at $\sim 250-350$ days. SNe~2015da and 2022mma do not show clear evidence for such an accelerated LC decline at least until $\sim 450$ days. This fact might indicate different dust formation, although this can also be due to the viewing angle effects or to the different radial distribution of the CSM. For more robust conclusions, we need proper observational campaigns including also infrared observations as in \citet[][]{Reynolds2025a}.

The profiles of the H$\alpha$ lines of the SNe after the broad component emerges ($t\gtrsim 100$ days) reflect the rough locations of the emitting regions in the SN ejecta. In cases with less dense CSM, the interaction shock can expand further out, and thus heat more outer parts of the SN ejecta. Therefore, fainter 21irp-like SNe would generally have wider H$\alpha$ lines. In fact, the fainter objects in the sample (SNe~2018khh, 2021irp, 2022oo and 2022mma) have relatively higher velocities, while the brighter ones (SNe~2010jl, 2015da and 2017hcc) have relatively lower velocities at $\sim 200$ days after the explosion (see Figure~\ref{fig:fig8}). Here, there is one exception though: SN~2021adxl is luminous and shows relatively high velocities though. At the same time, the shapes of the spectral lines originating from aspherical emitting regions should depend on the viewing angle as well. The consistency of the observed shapes of the H$\alpha$ lines with the disk CSM interaction scenario need to be investigated in more quantitatively, e.g., with radiation hydrodynamic simulations.

All the 21irp-like SNe with polarimetric observations (SNe~2010jl, 2015da, 2017hcc, 2021irp, 2021adxl, and 2022mma) show high polarization ($\sim1-6$ \%; Figure~\ref{fig:fig12}), and evolve similarly with a rapid decline from a polarization degree of several percents shortly after the explosion to a few percents at $\sim 50$ days from the LC peak followed by a relatively slow evolution to around $\sim1$ \% later on. The lack of wavelength dependence of the polarization suggests its origin to be an aspherical scattering-dominated photosphere created by the aspherical CSM interaction. The relatively similar polarization behavior in 21irp-like SNe suggests similar aspherical structures of the photosphere, and thus of their CSM. The viewing angle effects on the polarization degree might be seen in the small variation of the polarization degrees.

\subsection{Possible progenitor systems}

In this subsection, we speculate about the possible progenitor systems, based on the inferred CSM properties. We concluded that 21irp-like SNe have disk-like CSM whose masses ($\sim0.1-10$ M$_{\odot}$), radial distributions, and/or opening angles are diverse (see Section~\ref{sec:discussion_csm_properties}).
For the origin of SN~2021irp, \citet[][]{Reynolds2025b} proposed the possibility of a typical massive star like the progenitors of normal Type~II SNe in a binary system but with an extreme binary interaction, e.g., a common envelope evolution triggered by some unknown mechanism, just before the SN explosion. Since the observational properties, and thus the CSM properties of the other SNe in this sample are generally similar to those of SN~2021irp, they might also be explained with similar scenarios with slightly different binary and/or progenitor parameters.

The key information on these 21irp-like SNe is that they tend to be in the star-forming and generally low metallicity environments (see Section~\ref{sec:host}), although the metallicity of the host galaxy of SN~2018khh is slightly higher than solar metallicity. This is consistent with the above progenitor scenario, which needs a massive star with significant amount of hydrogen envelope. For example, in high metallicity environments, massive stars could easily lose their hydrogen envelopes via line-driven wind, and would not keep enough hydrogen envelope to create a nearby CSM inferred in the 21irp-like SNe. At the same time, the fraction of massive stars in close binary systems is increased at at low-metallicity environments \citep[e.g.,][]{Villasenor2025}, which might also be another reason for the environmental preference of the 21irp-like SNe. It is important to study the CSM properties more precisely and investigate the correlations between the CSM parameters (e.g., the CSM mass vs. opening angle, or the CSM mass vs. radial distribution), by increasing the sample size. This would further constrain the mass-ejection mechanism and the progenitor systems. We note that, since these systems are not spherically symmetric, the observational measurements (e.g., the luminosity evolution, the line widths, the polarization degrees, the extinction by newly formed dust), and thus the estimated ejecta and CSM properties should depend on the viewing angle. Therefore, it is important to statistically study these 21irp-like SNe, by increasing the sample size.

\subsection{Relations with the other types of interacting SNe}

We compared the SNe in our sample with the well-observed long-lived Type~IIn SNe, SNe~2010jl, 2015da and 2017hcc, and noticed the similarities in their photometric, spectroscopic and polarimetric properties. In particular, SN~2010jl is regarded as a prototypical long-lived Type~IIn SN that showed similar observational characteristics with other long-lived Type~IIn SNe. Therefore, also other long-lived Type~IIn SNe might essentially be similar systems as SN~2021irp.
In addition, since some Type II superluminous SNe with broad emission-dominated Balmer lines show similar photometric and spectroscopic properties \citep[e.g.,][]{Kangas2022,Pessi2025}, they might share similar radiation processes. Indeed, the host galaxies of Type II SLSNe are similar to those for the 21irp-like SNe \citep[faint dwarf galaxies with low metallicity][]{Inserra2018}.
We note that some Type II superluminous SNe are claimed to have too large total radiated energies to be powered by the CSM interaction of usual Type II SNe \citep[whose kinetic energy is $10^{51}$ erg; e.g.,][]{Kangas2022,Pessi2025} and they should have a different origin than SN~2021irp has. However, they might be explained by the viewing angle effects of such aspherical systems as 21irp-like SNe. If the CSM disk is very dense, and thus the CSM interaction is intensive in a 21irp-like system, a larger fractions of the generated radiation might escape to the CSM-free directions than to the CSM disk directions. This might explain the brightest Type II SLSNe and fainter 21irp-like SNe with the same system with different viewing angles (polar and equatorial directions, respectively). If this is true, we should find correlations between the brightness, the polarization degree, the shapes of the broad Balmer lines, and the P-Cygni profiles of the narrow Balmer lines. It is therefore important to carry out a similar analysis using polarimetry and LC modeling for Type II SLSNe.

\section{Conclusions} \label{sec:conclusion}

In this paper, we have presented results from our photometric, spectroscopic, and/or polarimetric observations of four 21irp-like events, SNe~2018khh, 2021adxl, 2022oo and 2022mma. Based on their observational properties and including SN~2021irp itself as well as well-observed bright and long-lived Type II SNe, SNe~2010jl, 2015da and 2017hcc, we have investigated their CSM characteristics. These events generally have luminous and long-lived LCs, with some variations in the brightness and the evolution (from $\sim -17$ to $\sim -20$ absolute mag in the $r$/$o$ band even at $\sim 200$ days after the explosion). They show photospheric spectra characterized mainly by broad Balmer emission lines for several hundreds of days, with some variations in the shapes of the lines. They show high polarization ($\sim1-3$ \% at the brightness peak) with rapid time evolution (from $\sim3-6$ \% before the peak to $\sim1$ \% at $\sim200$ days after the peak). The overall properties can qualitatively be explained by the disk-CSM-interaction scenario, that is, a typical Type~II SN interacting with disk-like CSM. The diversity in the observational properties suggest variations in the CSM properties, e.g., the CSM mass, the radial distribution, and/or the opening angle of the CSM disk. In the disk-CSM-interaction scenario, these variations in the CSM properties possibly originate from the difference of the binary parameters of the progenitor systems and/or of the properties of the progenitors and their companion stars.

\begin{acknowledgements}

This work is partly based on observations collected at the European Organisation for Astronomical Research in the Southern Hemisphere (ESO) under programme IDs 103.D-0338 (PI: Kuncarayakti), and as part of ePESSTO+ (the advanced Public ESO Spectroscopic Survey for Transient Objects Survey – PI: Inserra) and ePESSTO (PI: Smartt). ePESSTO+ observations were obtained under ESO program IDs 108.220C, while ePESSTO observations under ESO program IDs 199.D-0143.
This work is partly based on observations made under program IDs P63-016, P64-023 and P65-005 with the Nordic Optical Telescope, owned in collaboration by the University of Turku and Aarhus University, and operated jointly by Aarhus University, the University of Turku and the University of Oslo, representing Denmark, Finland and Norway, the University of Iceland and Stockholm University at the Observatorio del Roque de los Muchachos, La Palma, Spain, of the Instituto de Astrofisica de Canarias.
Based on observations obtained at the Southern Astrophysical Research (SOAR) telescope, which is a joint project of the Minist\'{e}rio da Ci\^{e}ncia, Tecnologia e Inova\c{c}\~{o}es (MCTI/LNA) do Brasil, the US National Science Foundation’s NOIRLab, the University of North Carolina at Chapel Hill (UNC), and Michigan State University (MSU).
We acknowledge ESA Gaia, DPAC and the Photometric Science Alerts Team (http://gsaweb.ast.cam.ac.uk/alerts).
This work has made use of data from the Asteroid Terrestrial-impact Last Alert System (ATLAS) project. The Asteroid Terrestrial-impact Last Alert System (ATLAS) project is primarily funded to search for near earth asteroids through NASA grants NN12AR55G, 80NSSC18K0284, and 80NSSC18K1575; byproducts of the NEO search include images and catalogs from the survey area. This work was partially funded by Kepler/K2 grant J1944/80NSSC19K0112 and HST GO-15889, and STFC grants ST/T000198/1 and ST/S006109/1. The ATLAS science products have been made possible through the contributions of the University of Hawaii Institute for Astronomy, the Queen’s University Belfast, the Space Telescope Science Institute, the South African Astronomical Observatory, and The Millennium Institute of Astrophysics (MAS), Chile.
Funding for the Sloan Digital Sky Survey (SDSS) has been provided by the Alfred P. Sloan Foundation, the Participating Institutions, the National Aeronautics and Space Administration, the National Science Foundation, the U.S. Department of Energy, the Japanese Monbukagakusho, and the Max Planck Society. The SDSS Web site is \url{http://www.sdss.org/}. The SDSS is managed by the Astrophysical Research Consortium (ARC) for the Participating Institutions. The Participating Institutions are The University of Chicago, Fermilab, the Institute for Advanced Study, the Japan Participation Group, The Johns Hopkins University, Los Alamos National Laboratory, the Max-Planck-Institute for Astronomy (MPIA), the Max-Planck-Institute for Astrophysics (MPA), New Mexico State University, University of Pittsburgh, Princeton University, the United States Naval Observatory, and the University of Washington.
T.N. and H.K. acknowledge support from the Research Council of Finland projects 324504, 328898 and 353019.
S.M. and T.M.R. acknowledge support from the Research Council of Finland project 350458. 
This work was funded by ANID, Millennium Science Initiative, ICN12\_009.
K.M. acknowledges support from JSPS KAKENHI grant (JP24H01810 and JP 24KK0070), and support from the JSPS Open Partnership Bilateral Joint Research Projects between Japan and Finland (JPJSBP120229923).
T.-W.C. acknowledges the Yushan Fellow Program by the Ministry of Education, Taiwan for the financial support (MOE-111-YSFMS-0008-001-P1).
T.E.M.B. is funded by Horizon Europe ERC grant no. 101125877.
M.G.B. acknowledges financial support from the Spanish Ministerio de Ciencia e Innovación (MCIN) and the Agencia Estatal de Investigación (AEI) 10.13039/501100011033 under the PID2023-151307NB-I00 SNNEXT project, from Centro Superior de Investigaciones Científicas (CSIC) under the PIE project 20215AT016 and the program Unidad de Excelencia María de Maeztu CEX2020-001058-M, and from the Departament de Recerca i Universitats de la Generalitat de Catalunya through the 2021-SGR-01270 grant.
M.K. acknowledges financial support from MICINN (Spain) through the programme Juan de la Cierva-Incorporación [JC2022-049447-I] and financial support from AGAUR, CSIC, MCIN and AEI 10.13039/501100011033 under projects PID2023-151307NB-00, PIE 20215AT016, CEX2020-001058-M, and 2021-SGR-01270.
T.P. acknowledges the financial support from the Slovenian Research Agency (grants I0-0033, P1-0031, J1-8136, J1-2460 and Z1-1853).
      
\end{acknowledgements}

  \bibliographystyle{aa} 
  \bibliography{aa.bib} 

\appendix

\section{Photometric observations for SN~2018khh} \label{sec:app_phot_18khh}

\begin{figure}
   \centering
            \includegraphics[width=\hsize]{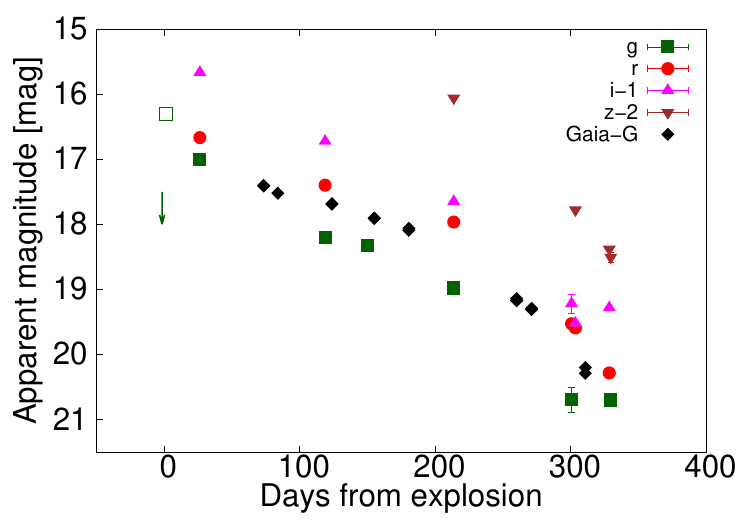}
      \caption{Light curves of SN~2018khh in the optical bands. The $G$-band LC obtained by Gaia \citep[][]{Gaia2016} is taken through the Gaia Photometric Science Alerts (\url{http://gsaweb.ast.cam.ac.uk/alerts/home}). The green open square and arrow symbols indicate the discovery and last non-detection $g$-band magnitudes, respectively \citep{Brimacombe2018}.
              }
      \label{fig:LC_18khh}
   \end{figure}

We obtained $g$-, $r$-, $i$-, and $z$-band photometry of SN~2018khh. The images of the SN were obtained with the Goodman instrument. The image reduction consisted in bias subtraction and flat fielding of the images and was performed with IRAF. The astrometric registration of the images was done with astrometry.net \citep[][]{Lang2010}, and the host galaxy light was subtracted from the SN images using HOTPANTS \citep[][]{Becker2015} employing template images obtained in 2022 when the SN had significantly faded. The photometry was computed performing a Point Spread Function (PSF) fit of the SN with python on host-galaxy-subtracted images. The results of photometry are provided in Table~\ref{tab:phot_18khh} and Figure~\ref{fig:LC_18khh}. The photometry of SN 2018khh was also presented in \citep[][]{Jacobson-Galan2024}. 

\begin{table*}
      \caption[]{Log and measurements of the polarimetric observations of SN 2021adxl.}
      \label{tab:phot_18khh}
      $
         \begin{array}{ccccc}
            \hline
            \noalign{\smallskip}
            \rm{MJD} & g & r & i & z\\
            (\rm{days}) & (\rm{mag}) & (\rm{mag}) & (\rm{mag}) & (\rm{mag})\\
            \noalign{\smallskip}
            \hline\hline
            \noalign{\smallskip}
            58497.04 & 17.00\pm0.01 & 16.67\pm0.01 & 16.66\pm0.02 & -\\
            \noalign{\smallskip}\hline \noalign{\smallskip}
            58589.39 & 18.20\pm0.01 & 17.40\pm0.01 & 17.72\pm0.01 & -\\
            \noalign{\smallskip}\hline \noalign{\smallskip}
            58620.39 & 18.33\pm0.03 & - &	- &	-\\
            \noalign{\smallskip}\hline \noalign{\smallskip}
            58684.31 & 18.98\pm0.02 & 17.96\pm0.02 & 18.65\pm0.02 & 18.05\pm0.02\\
            \noalign{\smallskip}\hline \noalign{\smallskip}
            58771.24 & 20.70\pm0.19 &	19.53\pm0.05 & 20.22\pm0.15 &-\\
            \noalign{\smallskip}\hline \noalign{\smallskip}
            58774.13 & - & 19.59\pm0.01 & 20.52\pm0.03 & 19.77\pm0.03\\
            \noalign{\smallskip}\hline \noalign{\smallskip}
            58799.10 & - & 20.28\pm0.03 & 20.28\pm0.03 & 20.38\pm0.03\\
            \noalign{\smallskip}\hline \noalign{\smallskip}
            58800.16 & 20.70\pm0.08 & - & - & 20.50\pm0.08\\
            \noalign{\smallskip}
            \hline
         \end{array}
         $
         \begin{minipage}{.88\hsize}
        \smallskip
        \end{minipage}
   \end{table*}

\section{Logs of the polarimetric and spectroscopic observations} \label{sec:app_obs}

\begin{table*}
      \caption[]{Log and measurements of the polarimetric observations of SN 2021adxl.}
      \label{tab:pol_21adxl}
      $
         \begin{array}{lccccccc}
            \hline
            \noalign{\smallskip}
            \rm{Date} & \rm{MJD} & \rm{Phase} & \rm{Exp. \;time} & \rm{Pol.\;degree} & \rm{Pol. \;angle} & \rm{Obs. \;mode} & \rm{Telescope}\\
            (\rm{UT}) & (\rm{days}) & (\rm{days}) & (\rm{seconds}) & (\%) & (\rm{degrees}) & & \\
            \noalign{\smallskip}
            \hline\hline
            \noalign{\smallskip}
            2022-03-01.35 & 59639.35 & +139.4 & 4 \times 150 & 1.19 \pm 0.05 & 10.0 \pm 1.5 & \rm{spec.} & \rm{VLT}\\
            \noalign{\smallskip}\hline \noalign{\smallskip}
            2022-03-26.07 & 59664.07 & +164.1 & 4 \times 300 & 1.18 \pm 0.04 & 9.3 \pm 1.1 & \rm{spec.} & \rm{VLT}\\
            \noalign{\smallskip}\hline \noalign{\smallskip}
            \multirow{2}{*}{2022-04-24.02} & \multirow{2}{*}{59693.02} & \multirow{2}{*}{+193.0} & 4 \times 60 & 0.89 \pm 0.10 & 0.4 \pm 3.1 & \rm{image (V)} & \multirow{2}{*}{\rm{NOT}}\\
            & & & 4 \times 50 & 1.13 \pm 0.08 & 15.2 \pm 1.9 & \rm{image (R)} & \\
            \noalign{\smallskip} \hline \noalign{\smallskip}
            \multirow{2}{*}{2022-06-03.91} & \multirow{2}{*}{59733.91} & \multirow{2}{*}{+233.9} & 4 \times 80 & 1.01 \pm 0.10 & 21.0 \pm 2.7 & \rm{image (V)} & \multirow{2}{*}{\rm{NOT}}\\
            & & & 4 \times 50 & 1.41 \pm 0.09 & 174.9 \pm 1.7 & \rm{image (R)} & \\
            \noalign{\smallskip}
            \hline
         \end{array}
         $
         
         \begin{minipage}{.88\hsize}
        \smallskip
        Notes. The phase is measured relative to the assumed explosion date (59500.00 MJD).
        \end{minipage}
   \end{table*}

\begin{table*}
      \caption[]{Log and measurements of the polarimetric observations of SN 2022mma.}
      \label{tab:pol_22mma}
      $
         \begin{array}{lccccccc}
            \hline
            \noalign{\smallskip}
            \rm{Date} & \rm{MJD} & \rm{Phase} & \rm{Exp. \;time} & \rm{Pol.\;degree} & \rm{Pol. \;angle} & \rm{Obs. \;mode} & \rm{Telescope}\\
            (\rm{UT}) & (\rm{days}) & (\rm{days}) & (\rm{seconds}) & (\%) & (\rm{degrees}) & & \\
            \noalign{\smallskip}
            \hline\hline
            \noalign{\smallskip}
            \multirow{2}{*}{2022-06-26.98} & \multirow{2}{*}{59756.98} & \multirow{2}{*}{+17.8} & 4 \times 100 & 3.10 \pm 0.10 & 54.7 \pm 1.0 & \rm{image (V)} & \multirow{2}{*}{\rm{NOT}}\\
            & & & 4 \times 100 & 3.67 \pm 0.10 & 49.6 \pm 0.8 & \rm{image (R)} & \\
            \noalign{\smallskip} \hline \noalign{\smallskip}
            \multirow{2}{*}{2022-07-20.94} & \multirow{2}{*}{59780.94} & \multirow{2}{*}{+41.7} & 4 \times 70 & 2.00 \pm 0.09 & 42.6 \pm 1.4 & \rm{image (V)} & \multirow{2}{*}{\rm{NOT}}\\
            & & & 4 \times 70 & 2.48 \pm 0.10 & 57.2 \pm 1.0 & \rm{image (R)} & \\
            \noalign{\smallskip} \hline \noalign{\smallskip}
            \multirow{2}{*}{2022-07-31.92} & \multirow{2}{*}{59791.92} & \multirow{2}{*}{+52.7} & 4 \times 70 & 1.82 \pm 0.11 & 46.0 \pm 1.7 & \rm{image (V)} & \multirow{2}{*}{\rm{NOT}}\\
            & & & 4 \times 70 & 1.52 \pm 0.10 & 48.7 \pm 1.9 & \rm{image (R)} & \\
            \noalign{\smallskip} \hline \noalign{\smallskip}
            \multirow{2}{*}{2022-08-22.87} & \multirow{2}{*}{59813.87} & \multirow{2}{*}{+74.6} & 4 \times 80 & 2.15 \pm 0.14 & 36.1 \pm 1.9 & \rm{image (V)} & \multirow{2}{*}{\rm{NOT}}\\
            & & & 4 \times 70 & 1.23 \pm 0.14 & 59.6 \pm 3.1 & \rm{image (R)} & \\
            \noalign{\smallskip} \hline \noalign{\smallskip}
            \multirow{2}{*}{2022-09-16.84} & \multirow{2}{*}{59838.84} & \multirow{2}{*}{+99.6} & 4 \times 80 & 3.10 \pm 0.42 & 173.9 \pm 3.8 & \rm{image (V)} & \multirow{2}{*}{\rm{NOT}}\\
            & & & 4 \times 80 & 1.64 \pm 0.26 & 22.4 \pm 4.4 & \rm{image (R)} & \\
            \noalign{\smallskip}\hline \noalign{\smallskip}
            2022-12-23.28 & 59936.28 & +224.1 & 4 \times 100 & 2.46 \pm 0.32 & 36.8 \pm 3.6 & \rm{image (R)} & \rm{NOT}\\
            \noalign{\smallskip}
            \hline
         \end{array}
         $
         
         \begin{minipage}{.88\hsize}
        \smallskip
        Notes. The phase is measured relative to the assumed explosion date (59739.23 MJD).        
        \end{minipage}
   \end{table*}

\begin{table*}
      \caption[]{Log of the spectroscopic observations of SN 2021adxl.}
      \label{tab:spec_21adxl}
      $
         \begin{array}{lcccc}
            \hline
            \noalign{\smallskip}
            \rm{Date} & \rm{MJD} & \rm{Phase} & \rm{Exp. \;time} & \rm{Instrument/Telescope}\\
            (\rm{UT}) & (\rm{days}) & (\rm{days}) & (\rm{seconds}) & \\
            \noalign{\smallskip}
            \hline\hline
            \noalign{\smallskip}
            2022-03-01.35 & 59639.35 & +139.4 & 600 & \rm{FORS2/VLT} \\
            \noalign{\smallskip} \hline \noalign{\smallskip}
            2022-03-26.07 & 59664.07 & +164.1 & 1200 & \rm{FORS2/VLT} \\
            \noalign{\smallskip} \hline \noalign{\smallskip}
            2022-04-24.01 & 59693.01 & +193.0 & 600 & \rm{ALFOSC/NOT} \\
            \noalign{\smallskip} \hline \noalign{\smallskip}
            2022-05-21.88 & 59720.88 & +220.9 & 600 & \rm{ALFOSC/NOT} \\
            \noalign{\smallskip} \hline \noalign{\smallskip}
            2022-06-05.91 & 59735.91 & +235.9 & 900 & \rm{ALFOSC/NOT} \\
            \noalign{\smallskip} \hline \noalign{\smallskip}
            2023-02-21.12 & 59996.12 & +496.1 & 1200 & \rm{ALFOSC/NOT} \\
            \noalign{\smallskip}
            \hline
         \end{array}
         $
         
         \begin{minipage}{.88\hsize}
        \smallskip
        Notes. The phase is measured relative to the assumed explosion date (59500.00 MJD).
        \end{minipage}
   \end{table*}

\begin{table*}
      \caption[]{Log of the spectroscopic observations of SN 2022mma.}
      \label{tab:spec_22mma}
      $
         \begin{array}{lcccc}
            \hline
            \noalign{\smallskip}
            \rm{Date} & \rm{MJD} & \rm{Phase} & \rm{Exp. \;time} & \rm{Instrument/Telescope}\\
            (\rm{UT}) & (\rm{days}) & (\rm{days}) & (\rm{seconds}) & \\
            \noalign{\smallskip}
            \hline\hline
            \noalign{\smallskip}
            2022-06-27.00 & 59757.00 & +17.8 & 600 & \rm{ALFOSC/NOT} \\
            \noalign{\smallskip} \hline \noalign{\smallskip}
            2022-07-13.91 & 59773.91 & +34.7 & 300 & \rm{ALFOSC/NOT} \\
            \noalign{\smallskip} \hline \noalign{\smallskip}
            2022-07-31.93 & 59791.93 & +52.7 & 300 & \rm{ALFOSC/NOT} \\
            \noalign{\smallskip} \hline \noalign{\smallskip}
            2022-08-22.88 & 59813.88 & +74.7 & 300 & \rm{ALFOSC/NOT} \\
            \noalign{\smallskip} \hline \noalign{\smallskip}
            2022-09-17.84 & 59839.84 & +100.6 & 300 & \rm{ALFOSC/NOT} \\
            \noalign{\smallskip} \hline \noalign{\smallskip}
            2023-01-24.18 & 59968.18 & +229.0 & 1200 & \rm{ALFOSC/NOT} \\
            \noalign{\smallskip} \hline \noalign{\smallskip}
            2023-02-21.16 & 59996.16 & +256.9 & 1200 & \rm{ALFOSC/NOT} \\
            \noalign{\smallskip} \hline \noalign{\smallskip}
            2023-03-30.04 & 60033.04 & +293.8 & 1200 & \rm{ALFOSC/NOT} \\
            \noalign{\smallskip}
            \hline
         \end{array}
         $
         
         \begin{minipage}{.88\hsize}
        \smallskip
        Notes. The phase is measured relative to the assumed explosion date (59739.23 MJD).
        \end{minipage}
   \end{table*}

\begin{table*}
      \caption[]{Log of the spectroscopic observations of SN 2022oo.}
      \label{tab:spec_22oo}
      $
         \begin{array}{lcccc}
            \hline
            \noalign{\smallskip}
            \rm{Date} & \rm{MJD} & \rm{Phase} & \rm{Exp. \;time} & \rm{Instrument/Telescope}\\
            (\rm{UT}) & (\rm{days}) & (\rm{days}) & (\rm{seconds}) & \\
            \noalign{\smallskip}
            \hline\hline
            \noalign{\smallskip}
            2022-01-25.30 & 59604.30 & +49.7 & 900 & \rm{EFOSC/NTT} \\
            \noalign{\smallskip} \hline \noalign{\smallskip}
            2022-04-23.98 & 59692.98 & +138.3 & 1800 & \rm{ALFOSC/NOT} \\
            \noalign{\smallskip} \hline \noalign{\smallskip}
            2022-05-24.24 & 59723.24 & +168.6 & 2700 & \rm{EFOSC/NTT} \\
            \noalign{\smallskip}
            \hline
         \end{array}
         $
         
         \begin{minipage}{.88\hsize}
        \smallskip
        Notes. The phase is measured relative to the assumed explosion date (59554.65 MJD).
        \end{minipage}
   \end{table*}

\begin{table*}
      \caption[]{Log of the spectroscopic observations of SN 2018khh.}
      \label{tab:spec_18khh}
      $
         \begin{array}{lcccc}
            \hline
            \noalign{\smallskip}
            \rm{Date} & \rm{MJD} & \rm{Phase} & \rm{Exp. \;time} & \rm{Instrument/Telescope}\\
            (\rm{UT}) & (\rm{days}) & (\rm{days}) & (\rm{seconds}) & \\
            \noalign{\smallskip}
            \hline\hline            
            \noalign{\smallskip} \hline \noalign{\smallskip}
            2018-12-21.09 & 58473.09 & +2.5 &  400 & \rm{Goodman\;HTS/SOAR} \\
            \noalign{\smallskip} \hline \noalign{\smallskip}
            2018-12-22.06 & 58474.06 & +3.5 &  1260 & \rm{Goodman\;HTS/SOAR} \\
            \noalign{\smallskip} \hline \noalign{\smallskip}
            2018-12-30.04 & 58482.04 & +11.5 & 3000 & \rm{EFOSC2/NTT} \\
            \noalign{\smallskip} \hline \noalign{\smallskip}
            2019-01-09.04 & 58492.04 & +21.5 & 1500 & \rm{EFOSC2/NTT} \\
            \noalign{\smallskip} \hline \noalign{\smallskip}
            2019-01-14.03 & 58497.03 & +26.5 & 1300 & \rm{Goodman\;HTS/SOAR} \\
            \noalign{\smallskip} \hline \noalign{\smallskip}
            2019-04-16.40 & 58589.40 & +118.8 & 1200 & \rm{Goodman\;HTS/SOAR} \\
            \noalign{\smallskip} \hline \noalign{\smallskip}
            2019-04-30.36 & 58603.36 & +132.8 & 1370 & \rm{FORS2/VLT} \\
            \noalign{\smallskip} \hline \noalign{\smallskip}
            2019-05-17.37 & 58620.37 & +149.8 & 1800 & \rm{Goodman\;HTS/SOAR} \\
            \noalign{\smallskip} \hline \noalign{\smallskip}
            2019-07-20.29 & 58684.29 & +213.7 & 2600 & \rm{Goodman\;HTS/SOAR} \\
            \noalign{\smallskip} \hline \noalign{\smallskip}
            2019-08-15.19 & 58710.19 & +239.6 & 5400 & \rm{Goodman\;HTS/SOAR} \\
            \noalign{\smallskip} \hline \noalign{\smallskip}
            2019-09-14.10 & 58740.10 & +269.6 & 2120 & \rm{Goodman\;HTS/SOAR} \\
            \noalign{\smallskip} \hline \noalign{\smallskip}
            2019-10-18.09 & 58774.09 & +303.5 & 3600 & \rm{Goodman\;HTS/SOAR} \\
            \noalign{\smallskip} \hline \noalign{\smallskip}
            2019-10-28.03 & 58784.03 & +313.5 & 5400 & \rm{EFOSC2/NTT} \\
            \noalign{\smallskip} \hline \noalign{\smallskip}
            2019-11-12.05 & 58799.05 & +328.5 & 6000 & \rm{Goodman\;HTS/SOAR} \\
            \noalign{\smallskip} \hline \noalign{\smallskip}
            2019-11-13.13 & 58800.13 & +329.6 & 2100 & \rm{Goodman\;HTS/SOAR} \\
            \noalign{\smallskip}
            \hline
         \end{array}
         $
         
         \begin{minipage}{.88\hsize}
        \smallskip
        Notes. The phase is measured relative to the assumed explosion date (58470.55 MJD).
        \end{minipage}
   \end{table*}

\end{document}